\documentclass[10pt]{article}
\usepackage{mathrsfs}
\usepackage[tbtags]{amsmath}
\usepackage{epsfig,amstext,amsthm,latexsym}
\usepackage{epstopdf}
\usepackage{amssymb,color}
\usepackage{rotating}
\usepackage{multirow}
\usepackage{relsize}
 \allowdisplaybreaks[4]
\usepackage{graphicx, subfigure}
\usepackage[T1]{fontenc}
\usepackage[utf8]{inputenc}
\usepackage{tabularx,ragged2e,booktabs,caption}

\newcolumntype{C}[1]{>{\Centering}m{#1}}

\definecolor{c20}{rgb}{0.,0.7,0.}
\definecolor{c30}{rgb}{0.,0.,1.}
\definecolor{c40}{rgb}{1,0.1,0.7}
\definecolor{c50}{rgb}{1,0,0}

\def\cL#1{\textcolor{c40}{#1}}
\def\cL#1{#1}

\catcode`\@=11 \@addtoreset{equation}{section} \catcode`\@=12

\allowdisplaybreaks[4] 
\newcommand{\nwc}{\newcommand}
\nwc{\COM}[1]{}

\nwc{\vs}[1]{\vskip #1 cm}

\newtheorem{theo}{Theorem}[section]
\newtheorem{sat}[theo]{Proposition}
\newtheorem{de}[theo]{Definition}
\newtheorem{lem}[theo]{Lemma}

\newtheorem{korr}[theo]{Corollary}
\newtheorem{remark}[theo]{Remark}
\newtheorem{exxa}[theo]{Example}

\newcommand{\nelem}[1]{{Lemma \ref{#1}}}
\newcommand{\neprop}[1]{{Proposition \ref{#1}}}
\newcommand{\netheo}[1]{{Theorem \ref{#1}}}
\newcommand{\nekorr}[1]{{Corollary \ref{#1}}}

\def\PF{q_t(s)}
\def\FRE{\mbox{Fr\'{e}chet}}

\def\I#1{\mathbb{I}\{#1\}}

\def\bar{\overline}
\def\tilde{\widetilde}

\newcommand{\ve}{\varepsilon}

\newcommand{\abs}[1]{\lvert #1 \rvert}
\newcommand{\Abs}[1]{ \Bigl \lvert #1 \Bigr \rvert}

\newcommand{\pk}[1]{{\mathbb{P}}\left\{#1\right\} }

\newcommand{\R}{\mathbb{R}}

\newcommand{\N}{\mathbb{N}}

\newcommand{\inn}{\in \N}

\newcommand{\limit}[1]{\lim_{#1 \to   \infty}}

\newcommand{\BQN}{\begin{eqnarray}}
\newcommand{\EQN}{\end{eqnarray}}
\newcommand{\BQNY}{\begin{eqnarray*}}
\newcommand{\EQNY}{\end{eqnarray*}}
\newcommand{\BS}{\begin{sat}}
\newcommand{\ES}{\end{sat}}
\newcommand{\BL}{\begin{lem}}
\newcommand{\EL}{\end{lem}}
\newcommand{\BT}{\begin{theo}}
\newcommand{\ET}{\end{theo}}
\newcommand{\BK}{\begin{korr}}
\newcommand{\EK}{\end{korr}}
\newcommand{\BD}{\begin{de}}
\newcommand{\ED}{\end{de}}
\newcommand{\BR}{\begin{remark}}
\newcommand{\ER}{\end{remark}}
\newcommand{\BIT}{\begin{itemize}}
\newcommand{\EIT}{\end{itemize}}
\newcommand{\BDI}{\begin{description}}
\newcommand{\EDI}{\end{description}}

\newcommand{\BEX}{\begin{exxa}}
\newcommand{\EEX}{\end{exxa}}

\newcommand{\QED}{\hfill $\Box$}
\newcommand{\IF}{\infty}

\def\fracl#1#2{\Bigr( \frac{#1}{#2} \Bigl) }

\newcommand{\prooftheo}[1]{ \textsc{Proof of Theorem} \ref{#1} }
\newcommand{\proofprop}[1]{\textsc{Proof of Proposition} \ref{#1}}
\newcommand{\prooflem}[1]{\textsc{Proof of Lemma} \ref{#1}}
\newcommand{\proofkorr}[1]{\textsc{Proof of Corollary} \ref{#1}}

\newcommand{\sign}{\mathrm{sign}}
\newcommand{\E}[1]{\mathbb{E}\left\{#1\right\}}
\newcommand{\expon}[1]{\exp\left(#1\right)}

\def\PF{q_t(s)}

\def\RV{\mathrm{RV}}
\def\ERV{\mathrm{ERV}}
\def\sign{\mathrm{sign}}
\def\c{\mathrm{c}}

\def\HG{Haezendonck-Goovaerts }
\def\VaR{{\rm VaR}}
\def\nesec#1{Section \ref{#1}}
\begin{document}
\title{ \bf  Approximations of Weyl fractional-order  integrals\\ with insurance applications\footnote{The authors would like to thank Enkelejd Hashorva  for useful discussions and suggestions}}
\author{Chengxiu Ling$^a$, Zuoxiang Peng$^b$ \\ 
\small{$^a$Department of Actuarial Science, University of Lausanne,} \\
\small{UNIL-Dorigny, 1015 Lausanne, Switzerland}\\
\small{$^b$School of Mathematics and Statistics, Southwest   University, 400715 Chongqing, China}
}

       \maketitle
 {\bf Abstract:}
In this paper, we investigate the approximations of generalized Weyl fractional-order integrals in extreme value theory framework. We present three applications of our asymptotic results concerning the higher-order tail approximations of  deflated risks as well as approximations of \HG and expectile risk measures. Illustration of the obtained results is done by various examples and some numerical analysis.



\section{Introduction}\label{sec1}
\COM{Given a constant ${\kappa>-1}$ and a measurable weight function $w(\cdot)$ on $(0,\IF)$, denote by 
$$\mathcal D_{\kappa,w}=\{F: F\mbox{\ a\ distribution\ function\ with\ $ {F(0)=0}$} \ \mbox{such\ that\  $\int_\epsilon^\IF x^{\kappa}|w(x)|\, dF(x)<\IF,\ \forall \epsilon>0$}\}.
$$ 
If a random risk $X\sim F\in\mathcal D_{\kappa,w}$, then the Weyl-Stieltjes fractional-order integral operator $\mathcal J_{\kappa, w}$ is defined on $F$ as follows
\BQNY
({\mathcal J_{\kappa, w}}
F)(x)&:=&{\frac1{\Gamma(\kappa+1)}\int_x^\IF w(y)(y-x)^\kappa
\,dF(y)} \\
&=&\frac1{\Gamma(\kappa+1)}\E{w(X) {(X-x)_+^\kappa}},\quad x>0.
\EQNY
The integral operator ${\mathcal J_{\kappa, w}}$ 
is extremely useful in both the theoretical study concerning for instance the tail asymptotics under beta random scaling in \cite{HashorvaP2010}, and in some statistical applications such as the Wicksell problem in \cite{ReissT2007}. Indeed, the Weyl fractional-order integral 
$\mathcal J_{\kappa, w}F$ has an alternative form related to Beta random scaling: for a (independent of $X$) scaling factor $S\sim Beta(1, \kappa), \kappa>0$, we have 
\BQNY
({\mathcal J_{\kappa , w}}
F)(x) = \frac{ \E{X^\kappa w(X) \I{SX>x}}}{\Gamma(\kappa +1)}. 
\EQNY
}
Throughout this paper, let $X$ be a random risk with distribution function (df) $F$ (denoted by $X\sim F$), and $S\in(0,1)$ an independent of random risk. Of interest is the following integral $\mathcal I_{\mathcal L, S}(x, X)$ given by
\BQNY
\mathcal I_{\mathcal L, S}(x,X) := \E{\mathcal L(X) \I{SX>x}}
\EQNY
for some given function $\mathcal L(\cdot)$ such that the integral is well-defined. Here $\I{\cdot}$ stands for the indicator function. The integral $\mathcal I_{\mathcal L, S}(x,X)$ is closely related to the Weyl fractional-order integral;
see e.g., \cite{HashorvaP2010, PakesN2007, ReissT2007} for related applications on beta random scaling and Wicksell problem. 
\\
In various theoretical and practical situations, the question arising naturally is how the approximation of $\mathcal I_{\mathcal L, S}(x,X)$ as $x$ goes to  
 the right-endpoint of $F$, is influenced by the tail behavior of $S$ and $X$. For instance  \cite{HashorvaLP14, HashorvaPT2010} studied the asymptotics of the tail of deflated risk $SX$ which is reduced by $\mathcal I_{\mathcal L, S}(x,X)$ with $\mathcal L(x)\equiv1$. \\
Another motivation for considering the approximations of $\mathcal I_{\mathcal L, S}(x,X)$ comes from finance and risk management fields. Particularly, if $S$ is a Beta distributed random variable with parameters 1 and $\kappa, \kappa>0$ (denoted by $X\sim Beta(1, \kappa)$),  and $\mathcal L(x)= x^\kappa$, then, with $x_+=\max(x,0)$ 
\BQNY
\mathcal I_{\mathcal L, S}(x,X) = \E{(X-x)_+^\kappa},
\EQNY
which is closely related to several risk measures such as  the \HG (H-G) and expectile risk measures; see \cite{BelliniB2014, BelliniRosazza2012, BelliniKMR2014, MaoH2012, MaoHN2015, TangY2012}, and references therein for related discussions. 

In this paper, we are interested in the derivation of some approximations of the integral   $\mathcal I_{\mathcal L, S}(x,X)$ with $\mathcal L(x)= x^\kappa$ for some given constant $\kappa$, abbreviate it as $\mathcal I_{\kappa, S}(x,X)$, i.e.,  
\BQN\label{def:GWI}
\mathcal I_{\kappa, S}(x,X) := \E{X^\kappa \I{SX>x}}. 
\EQN
We remark that one could similarly consider the general function $\mathcal L(\cdot)$  by applying
the methodology of regular variations; for some technical reasons we study only the power function. 

Our principle results,  \netheo{T1} and \netheo{T2}, are concerned with the second- and third-order approximations of \eqref{def:GWI}.  
Our main methodology is based on the higher-order regular variation theory which was discussed deeply by \cite{deHaanS1996, FragaAlvesdL2006,  WangC2006}. 
As a by-product, we establish a technical inequality, namely the extensional Drees\rq{} inequality for third-order regular variations (see \nelem{L1} in the Appendix), which is of its own interest; see e.g., \cite{deHaanF2006, DraismadPP1999, FragaAlvesdL2006, MaoBook2013, Neves2009, WangC2006} for related discussions. \\
As important applications of these results, we shall discuss the tail asymptotics of deflated risks in \netheo{T5}, and \netheo{T6} (which deal with the cases that $X$ has Weibull and Gumbel tails in a unified way), refining those in  \cite{HashorvaLP14,HashorvaPT2010} where they only studied the first- and second-order tail approximations of $SX$ under three different tail behavior of $S$ and $X$.
Moreover, with the aid of \nekorr{Cor1}, dealing with the higher-order approximations of  \eqref{def:GWI} for $S\sim Beta(1, \kappa)$, we investigate 
 the approximations of  H-G and expectile risk measures which were initially studied by  \cite{TangY2012} and  \cite{BelliniB2014}, respectively. The two competitive risk measures have received more and more attention due to their nice mathematical and statistics properties; see e.g.,  \cite{BelliniKMR2014, CaiW2014, TangY2014}. 
As expected, our \netheo{T3} and \netheo{T4} are significant refinements of findings displayed in \cite{MaoH2012, TangY2012} and  \cite{BelliniB2014, MaoHN2015}. 

\COM{Specifically, if  $X$ is  a risk variable such that 
$\E{X_+^\kappa }<\IF, \kappa\ge1$ and $\pk{X=x_F}=0$, then the H-G risk measure for $X$ at level $q\in(0,1)$ is given by (set below 
$x_+=\max(0, x), \ x_-=\min(0, x)$)
\BQN\label{def: HG}
H_q[X]=x+\fracl{\E{(X-x)_+^\kappa }}{1-q}^{1/\kappa},
\EQN
where $x=x(q)\in(-\IF, x_F)$ is the unique solution to the equation
\BQN\label{Eq: kq}
\frac{(\E{(X-x)_+^{\kappa-1}})^\kappa }{(\E{(X-x)_+^\kappa })^{\kappa-1}} = 1-q,\quad \kappa>1
\EQN
and $x= \VaR_q(X) (:= F^\leftarrow(q))$ for $\kappa=1$.  Further if $X$ has finite expectation, then the expectile for $X$ at level $q\in(0,1)$  is defined as the unique minimizer of an
asymmetric quadratic loss function as follows 
\BQNY e_q= \arg \min_{x\in\R}
\big(q\E{(X-x)_+^2} +(1-q)\E{(X-x)_-^2}\big)
\EQNY
or equivalently as the unique solution of the equation 
\BQN\label{def: expectile}
e_q   = \E{X}+ \frac{2q-1}{1-q}\E{(X-e_q)_+}. \EQN
Since from \cite{TangY2012, BelliniB2014} that both $H_q[X]$ and $e_q$ tend to $x_F$, the right endpoint of $F$, as $q$ goes to 1, we see that our \nekorr{Cor1} is of crucial importance to derive the approximations of them; see \netheo{T3} and \netheo{T4}. 
As expected, our results are significant refinements of findings displayed in \cite{MaoH2012, TangY2012} and  \cite{BelliniB2014, MaoH2015}.  \\ 
It is worth mentioning that we only consider  the approximations of the H-G and expectile risk measures for the risk variable $X$ having  \FRE\ and Weibull tails. We believe that one can follow the same idea for those $X$ with Weibull-type tails; see for related discussions \cite{}. 

Expectiles are becoming quite popular in the risk management literature since it is the only coherent risk
measure possessing elicitability, i.e., it can be defined as the
minimizer of a suitable expected scoring function; see e.g.,
\cite{BelliniB2014, BelliniKMR2014, Gneiting2011}. In financial applications
 expectiles are alternatives to Value-at-Risk and to Expected
Shortfall (ES) risk measures. We refer to \cite{BelliniB2014, BelliniKMR2014,
HuaJ2011} for the discussions on expectiles and ES.
The contribution \cite{BelliniB2014} studied the asymptotic behavior of expectiles $e_q$ as $q\uparrow1$
 by imposing certain tail conditions on the risk variable $X$ from extreme  value theory framework.
}

The rest of this paper is organized as follows. In \nesec{sec2}, we establish our main results
following by the three applications including the approximations of deflated risks as well as  higher-order approximations of H-G and expectile risk measures. 
\nesec{sec4} is devoted to 
several illustrated examples and  some numerical analysis. 
All the proofs are relegated to \nesec{sec5}. We conclude this paper with an Appendix containing a technical inequality.

\section{Main Results}\label{sec2}
We start with the definitions and some properties of  regular variations which are key to establish our main results.

A measurable function $f: {[0, \IF)} \rightarrow \R$ is said to be an extended regularly varying  function (ERV) at infinity with index
$\gamma\in \R$, denoted by
$f\in \ERV_\gamma$, if (cf. \cite{BinghamGT1987, deHaanF2006})
\BQN\label{def:ERV}
\lim_{t\to\IF}\frac{f(tx)-f(t)}{a(t)}=\frac{x^\gamma-1}{\gamma} :=D_\gamma(x)
\EQN
holds for all $x>0$ and an eventually positive function $a(\cdot)$, which is referred to as the auxiliary function. 
In the meanwhile, $f$ is  regularly varying with index $\gamma$, denoted by $f\in \RV_\gamma$, if the limit in \eqref{def:ERV} holds with $(f(tx)-f(t))/a(t)$ and $D_\gamma(x)$ replaced by $f(tx)/f(t)$ and $x^\gamma$, respectively. 
\\
$\ERV$ and $\RV$ are powerful tools in the study of extreme value of statistics since it provides a suitable framework to study key features and properties of dfs belonging to max-domains of attractions. Namely, a df $F$ is said to be in the max-domain attraction (MDA) of $G_\gamma(x):=\expon{-(1+\gamma x)_+^{-1/\gamma}}, \gamma\in\R$, 
 i.e., there exist some constants $a_n>0, b_n\in\R$ such that
\BQNY
 \lim_{n\to\IF} \sup_{x\in\R}\abs{ F^n(a_n x+ b_n) - G_\gamma(x)} =0,
\EQNY
which holds if and only if $U\in \ERV_\gamma$ with $U(t):=F^\leftarrow(1-1/t)$ the tail quantile function; see \cite{deHaanF2006, EmbrechtsKM1997}. 
The df $F$ is so-called in  the \FRE, Gumbel and Weibull MDA according to $\gamma>0,  \gamma=0$ and $\gamma <0$, respectively. 
\\
 In this paper, we mainly use the second-  and third-order $\ERV$ and $\RV$ extensions, which are mainly used to investigate the speed of convergence of the first-  and second-order expansions of certain quantities of interest in different contexts; see e.g., \cite{CaideHZ2013,  deHaanP1997, LiPX2011, MaoH2012, OliveriraGF2006, PengNL2010}. 
\\
Refining \eqref{def:ERV}, we say that $f$ is of second-order extended regular
variation with parameters $\gamma\in \R$ and $\rho\le0$, denoted by
$f\in 2\ERV_{\gamma, \rho}$, if there exist some auxiliary functions $a(\cdot)$ eventually positive, and $A(\cdot)$ with
constant sign {near} infinity satisfying
$\lim_{t\rightarrow\infty} A(t)=0$, such that for all $x>0$ (cf. \cite{deHaanS1996, Resnick2007})
\begin{equation}\label{def1}
\lim_{t\rightarrow\infty}\frac{(f(tx)-f(t))/a(t)-D_\gamma(x)}{A(t)}=\int_1^x y^{\gamma-1}\int_1^y u^{\rho-1}\,du\,dy :=H_{\gamma, \rho}(x).
\end{equation}
Further, we shall write $f\in3\ERV_{\gamma, \rho, \eta}$ meaning that $f$ is of third-order regular variation
with parameters $\gamma\in\R$ and $\rho, \eta\le 0$, if there exist first-, second- and third-order auxiliary functions $a(\cdot)$ eventually positive, and $A(\cdot), B(\cdot)$
 with constant sign near infinity satisfying  $\lim_{t\to\IF}A(t)=\lim_{t\to\IF}B(t)=0$, such that for all $x>0$ (cf. \cite{FragaAlvesdL2006,WangC2006})
\BQN\label{ThirdLimit}
\limit t\frac{\frac{(f(tx)-f(t))/a(t) -D_\gamma(x) }{A(t)} -H_{\gamma,
\rho}(x)}{B(t)}=\int_1^x y^{\gamma-1}\int_1^y u^{\rho-1}\int_1^u v^{\eta-1}\,dv\,du\,dy := R_{\gamma, \rho, \eta}(x). 
\EQN
Similarly, we define $f\in 2\RV$ (or $f\in 3\RV$) with auxiliary function $A(\cdot)$ (and $B(\cdot)$) if  the limit in \eqref{def1} (or \eqref{ThirdLimit}) holds with $(f(tx)-f(t))/a(t)$ and $D_\gamma(x), H_{\gamma,\rho}(x)$ (and $R_{\gamma, \rho, \eta}(x)$) replaced by $f(tx)/f(t)$ and $x^\gamma, x^\gamma D_\rho(x)$ (and $x^\gamma D_{\rho+\eta}(x)$), respectively. \\
Note in passing that 
for $f\in 3\RV_{\gamma, \rho, \eta}$ with auxiliary functions $A(\cdot)$ and $B(\cdot)$, we have  from \cite{FragaAlvesdL2006} that $A\in2\RV_{\rho, \eta}$ with auxiliary function $B$, and $|B| \in\RV_\eta$. 


Throughout this paper, we write $\bar Q:= 1-Q$ for some function $Q$ and $U(t):= F^\leftarrow(1-1/t), t\ge1$ for the tail quantile function of $X$. 
By $\Gamma(\cdot)$ and $B(\cdot, \cdot)$ we mean the Euler Gamma function and Beta function. 
All the limits are taken as the argument goes to $x_F=U(\IF)$, the right endpoint of $X$ unless otherwise stated.

Our first result, \netheo{T1}, investigates the approximations of $\mathcal{I}_{\kappa,S}(x,X)$ given by \eqref{def:GWI} for $X$ being in the  \FRE\ MDA. 
Further, for $\kappa<\alpha, \varrho, \varsigma\le0$, we denote  $d_{0,\kappa}=\E{S^{\alpha-\kappa}}$ and 
\BQNY
&&d_{1,\kappa}=\frac{\E{S^{\alpha-\kappa-\varrho}}- \E{S^{\alpha-\kappa}}}{\varrho}
+\frac{\kappa \E{S^{\alpha-\kappa-\varrho}}}{\alpha(\alpha-\kappa-\varrho)},\quad  d_{2,\kappa}=\frac{\kappa(\E{S^{\alpha-\kappa-2\varrho}}- \E{S^{\alpha-\kappa-\varrho}})}{\alpha\varrho(\alpha-\kappa-\varrho)}\\
&& d_{3,\kappa}=\frac{\kappa}{\alpha}\left(\frac{\E{S^{\alpha-\kappa-\varrho-\varsigma}}- \E{S^{\alpha-\kappa-\varrho}}}{(\alpha-\kappa-\varrho)\varsigma}+\frac{ \E{S^{\alpha-\kappa-\varrho-\varsigma}}}{\alpha-\kappa-\varrho-\varsigma}\right) +\frac{\E{S^{\alpha-\kappa-\varrho-\varsigma}}- \E{S^{\alpha-\kappa}}}{\varrho+\varsigma}.
\EQNY
\BT\label{T1} 
(i) If $\bar F\in 2\RV_{-\alpha, \varrho}$ with auxiliary function $ A $ for some $\alpha>0$ and $\varrho\le 0$, 
then, for  $\mathcal I_{\kappa, S}(x,X)$  given by \eqref{def:GWI} with $\kappa<\alpha$
\BQN\label{Expan.Second}
\frac{\mathcal{I}_{\kappa,S}(x,X)}{ x^{\kappa}\bar F(x)}=  \frac{\alpha}{\alpha-\kappa}\left(d_{0,\kappa} + d_{1,\kappa}A (x)(1+o(1))\right).
\EQN
(ii) If $\bar F\in 3\RV_{-\alpha, \varrho, \varsigma}$ with auxiliary functions $ A $ and $ B $, then 
\BQN\label{Expan.Third}
\frac{\mathcal{I}_{\kappa,S}(x,X)}{x^{\kappa}\bar F(x)}=  \frac{\alpha }{\alpha-\kappa}\left(d_{0,\kappa} + A (x)\big(d_{1,\kappa} + d_{2,\kappa}A(x)(1+o(1)) + d_{3,\kappa} B (x)(1+o(1))\big)\right).
\EQN
\ET
{\remark (i) Recalling that the Weyl fractional-order integral $\mathcal{J}_{\beta+1,K_c} (x, X)$ with weight function $K_c(x):= x^c$ is given by (cf. \cite{HashorvaP2010}) 
\BQNY
\mathcal{J}_{\beta+1,K_c} (x, X)= \frac{\mathcal{I}_{\kappa,S}(x,X)}{\Gamma(\beta+1)} , \quad {\rm with\ } S\sim Beta(1, \beta),\  \kappa=\beta+c,
\EQNY
an immediate application of \eqref{Expan.Second} and \eqref{Expan.Third} together with 
$\E{S^l}=lB(l+1, \beta+1), \ l>0$ 
implies the second- and third-order expansions of $\mathcal{J}_{\beta+1,K_c} (x, X)$ extending Theorem  7.2 in \cite{HashorvaP2010}. 
\COM{\BQNY
\lefteqn{\mathcal{J}_{\beta+1,K_c} (x, X) = \frac{\alpha}{\alpha-\kappa} x^\kappa \bar F(x)\left( \frac{\Gamma(\alpha+1-\kappa)}{\Gamma(\alpha+1-c)} \right.}\\
&&\left.+ \frac{ A (x)}{\varrho}\left(\frac{\Gamma(\alpha+1-\varrho-\kappa)}{\Gamma(\alpha+1-\varrho-c)} -
\frac{\Gamma(\alpha+1-\kappa)}{\Gamma(\alpha+1-c)} +\frac{\kappa\varrho}{\alpha}\frac{\Gamma(\alpha-\varrho-\kappa)}{\Gamma(\alpha-\varrho-c+1)} \right) (1+o(1))\right)
\EQNY
which is the second-order expansion of $\mathcal{J}_{\beta+1,K_c}$ generalizing Theorem  7.2 in \cite{HashorvaP2010}. Particularly, with $c=0$ we have
\BQNY
\lefteqn{\E{(X-x)_+^\kappa}= \frac{\alpha\Gamma (\kappa+1)}{\alpha-\kappa} x^\kappa
 \bar F(x)\left( \frac{\Gamma(\alpha+1-\kappa)}{\Gamma(\alpha+1)} \right.}\\
&&\left.+ \frac{ A (x)}{\varrho}\left(\frac{\Gamma(\alpha+1-\varrho-\kappa)}{\Gamma(\alpha+1-\varrho)} -
\frac{\Gamma(\alpha+1-\kappa)}{\Gamma(\alpha+1)} +\frac{\kappa\varrho}{\alpha}\frac{\Gamma(\alpha-\varrho-\kappa)}{\Gamma(\alpha-\varrho+1)} \right) (1+o(1))\right).
\EQNY
One can also obtain the third-order expansions of $\mathcal{J}_{\beta+1,K_c}$ by alternatively using  \eqref{Expan.Third}. } 
\\
(ii) We see that the speed of convergence of the second-order expansion is determined by the two auxiliary functions $A$ and $B$, i.e., the parameters $\rho$ and $\varrho$. Most common risks are in the third-order Hall class defined by \eqref{def:ThirdHall} below, i.e., satisfy the third-order regularly varying conditions with equal  $\varrho$ and $\varsigma$; see \cite{CaeiroG2008,CaideHZ2013}.
}
\\
Next, we consider that $X$ is in the Gumbel and Weibull MDA.  
Hereafter, denote below for  $\gamma\in\R, \alpha>0, \rho, \eta\le 0, \varrho<0$ with $c_{\alpha, l}=\alpha(\alpha-1)\cdots (\alpha-l+1)/l!, l\inn$ and $D_\gamma, H_{\gamma, \rho}, R_{\gamma, \rho, \eta}$ given by \eqref{ThirdLimit} 
\BQN\label{def: notaion}
\left\{ \begin{array}{l}
L_{\alpha}=\int_0^1(D_\gamma(1/s))^\alpha\,ds,\quad M_{\alpha, l}=c_{\alpha,l}\int_0^1(D_\gamma(1/s))^{\alpha-l} (H_{\gamma, \rho}(1/s))^l\,ds\\
N_{\alpha,l, \varrho}=c_{\alpha, l}\int_0^1(D_\gamma(1/s))^{\alpha-l} (H_{\gamma, \rho}(1/s))^l\frac{(D_\gamma(1/s))^{-\varrho}-1}{\varrho}\,ds\\
 Q_{\alpha}=\alpha\int_0^1(D_\gamma(1/s))^{\alpha-1} R_{\gamma, \rho, \eta}(1/s)\,ds.
 \end{array}
 \right.
\EQN

\BT\label{T2} (i) If $U\in 2\ERV_{\gamma, \rho}$ with auxiliary functions $a,  A $ for some $\gamma, \rho\le 0$, and $\bar G(1-1/x)\in 2\RV_{-\alpha , \varrho }, \alpha>0, \varrho<0$ with auxiliary function $\tilde A$, then 
\BQN\label{Expan.Second.GW}
\frac{\mathcal{I}_{\kappa,S}(x,X)}
{x^\kappa  \bar F(x) \bar G\left(1-1/{\varphi_t}\right)} = L_{\alpha } +\left(M_{\alpha,1}  A (t) +N_{\alpha, 0, \varrho }\tilde A(\varphi_t) +(\kappa-\alpha ) \frac{L_{\alpha +1}}{\varphi_t}\right)(1+o(1)).
\EQN
(ii) If  $U\in 3\ERV_{\gamma, \rho, \eta}$ with auxiliary functions $a,  A $ and $ B $ for some $\gamma, \rho, \eta\le 0$, and  $\bar G(1-1/x)\in 3\RV_{-\alpha , \varrho , \varsigma }, \alpha>0, \varrho,  \varsigma<0$ with auxiliary functions $\tilde A$ and $\tilde B$, then
\BQN\label{Expan.Third.GW}
\lefteqn{\frac{\mathcal{I}_{\kappa,S}(x,X)}
{ x^\kappa \bar F(x) \bar G\left(1-1/{\varphi_t}\right)} =L_{\alpha } + M_{\alpha,1}  A (t) +N_{\alpha, 0, \varrho }\tilde A(\varphi_t)+(\kappa-\alpha ) \frac{L_{\alpha +1}}{\varphi_t} }\notag\\
&& +\left(A (t) \Big(M_{\alpha,2}  A (t)+ Q_{\alpha}  B (t) + N_{\alpha,1, \varrho } \tilde A(\varphi_t) +  (\kappa-\alpha)\frac{M_{\alpha+1,1}}{\varphi_t}\Big)+ (\kappa-\alpha)(\kappa-\alpha -1)\frac{L_{\alpha +2}}{2\varphi_t^2}\right.\notag \\
 &&\left.+\tilde A(\varphi_t)  \left(N_{\alpha , 0, \varrho +\varsigma }\tilde B(\varphi_t) + \Big((\kappa-\alpha ) N_{\alpha +1,0, \varrho } +L_{\alpha -\varrho +1} \Big) \frac{1}{\varphi_t}\right) \right)(1+o(1))
\EQN
with $\varphi_t=U(t)/a(t), t=1/\bar F(x)$ and $L_\alpha, M_{\alpha, l}, N_{\alpha, l,\varrho}, Q_{\alpha}$ given by \eqref{def: notaion}.
\ET
{\remark (i) It is possible to allow  $\varrho, \varsigma$ to be non-positive for $\gamma<0$. 
\\
(ii) We see that \netheo{T2} 
conducts a unified way in terms of the tail quantile function. Further, the speed of convergence seems more involved in the related parameters and auxiliary functions.
}

Next, we specify the expansions of $\mathcal I_{\kappa, S}(x, X)$ where we consider $S\sim Beta(1, \kappa)$ and certain third-order regularly varying conditions are imposed on $X$ (which is helpful to calculate explicitly the coefficients involved). In the meanwhile, 
 we will see that \nekorr{Cor1} given below is of crucial importance in the derivation of  the tail asymptotics of H-G and expectile  risk measures; see Sections \ref{Appl.HG} and \ref{Appl.Expectiles}.

In what follows, set  for $\gamma\neq0, \rho\le 0$
  \begin{gather}
  \begin{aligned} \label{def: xi}
&\xi_{\kappa ,\rho}=B\left(\frac{1-\rho}\gamma-\kappa,\kappa\right)\I{\gamma>0}+B\left(1-\frac{1-\rho}\gamma, \kappa\right)\I{\gamma<0} \\
&\tilde M_{\kappa,1}= \frac1{\gamma\rho}\left(\frac{\xi_{\kappa ,\rho}}{\xi_{\kappa ,0}} - \frac{\xi_{\kappa -1,\rho}}{\xi_{\kappa -1,0}}\right),\quad \Delta_\kappa =\frac1{\gamma\rho}\left(\kappa\frac{\xi_{\kappa -1,\rho}}{\xi_{\kappa -1,0}}
- (\kappa-1)\frac{\xi_{\kappa ,\rho}}{\xi_{\kappa ,0}}-1\right).
\end{aligned}
\end{gather}
\BK\label{Cor1} 
If (i) $U\in3\RV_{\gamma, \rho,\eta}$ with auxiliary functions $A$ and $B$ with $\gamma>0, \rho, \eta\le 0$,
or  (ii) $x_F-U\in3\RV_{\gamma,\rho,\eta}$ with $\gamma<0, \rho, \eta\le0$ and auxiliary functions $A $ and $B $,
 then  for all $\kappa>0$ such that $\kappa\gamma<1$  
\BQN\label{expans}
\frac{\E{(X-U(t))_+^\kappa }}{t^{-1}(a(t))^\kappa } = L_\kappa  +M_{\kappa,1}A(t)  + M_{\kappa ,2}A^2(t) (1+o(1)) + Q_\kappa A(t)B(t)(1+o(1)), \quad t\to\IF
\EQN
 holds with  $a(t)=\gamma U(t)\I{\gamma>0} -\gamma (x_F-U(t))\I{\gamma<0}, L_\kappa = \kappa\xi_{\kappa ,0}/{|\gamma|^\kappa }$ and
\BQNY
&&M_{\kappa,1} =\frac{\kappa\sign(\gamma)}{|\gamma|^{\kappa+1}\rho}\left( \xi_{\kappa ,\rho}-\xi_{\kappa ,0}\right), \ Q_\kappa =\frac{\kappa\sign(\gamma)}{|\gamma|^{\kappa+1}(\rho+\eta)}\left( \xi_{\kappa ,\rho+\eta}-\xi_{\kappa ,0}\right). \\
&& M_{\kappa ,2}=\frac{\kappa}{2|\gamma|^{\kappa+2}\rho^2}
\big((1-2\rho-\gamma)\xi_{\kappa ,2\rho} -2(1-\rho-\gamma)\xi_{\kappa ,\rho}+(1-\gamma)\xi_{\kappa ,0}\big).
\EQNY
\EK
{\remark \label{rem2} Note that the coefficients in \eqref{expans} are understood as their limits when $\rho$ or $\eta$ are zeros. Specifically, 
we have, with $l=1,2$
\BQNY
\lim_{\rho\to0}M_{\kappa,l} =  \frac{c_{\kappa,l} \sign(\gamma)}{{\abs\gamma}^{\kappa+l}}\omega_{\kappa,l},\quad \lim_{\eta\to0}Q_\kappa = \frac{\kappa \sign(\gamma)}{{\abs\gamma}^{\kappa+1}}
\left(\omega_{\kappa ,1}\I{\rho\neq0}+\frac{\widetilde\omega_{\kappa}}{2\abs\gamma}\I{\rho=0}\right),
\EQNY
where
\BQNY
&&\omega_{\kappa ,l}=\frac1{\abs\gamma}\int_0^\IF x^l(1-e^{-x})^{\kappa-l}\exp\left(-\frac{1-\kappa\gamma\I{\gamma>0}-\gamma \I{\gamma<0}}{\abs{\gamma}}x\right)\,dx\\
&&\widetilde\omega_{\kappa}=\frac1{\abs\gamma}\int_0^\IF x^{2}(1-e^{-x})^{\kappa-1}\exp\left(-\frac{1-\kappa\gamma\I{\gamma>0}-\gamma \I{\gamma<0}}{\abs{\gamma}}x\right)\,dx.
\EQNY
}
\section{Applications}\label{sec3}
In this section, we present three applications in insurance fields, namely the higher-order tail expansions of the deflated risks which refine those in \cite{HashorvaLP14}, and approximations of H-G  and expectile risk measures.
\subsection{Asymptotic expansions of deflated risks}
In the following, we apply \netheo{T1} and \netheo{T2} with $\kappa=0$ to obtain the third-order expansions of the tail of deflated risk $SX$ refining those in \cite{HashorvaLP14,HashorvaPT2010}. 
\BT\label{T5} 
Under the conditions as in \netheo{T1} (ii), we have
\BQNY
\frac{\pk{SX >x} }{\bar F(x)}=\E{S^\alpha} + \frac{\E{S^{\alpha-\varrho}} -\E{S^{\alpha}}}{\varrho} A(x) +  \frac{\E{S^{\alpha-\varrho-\varsigma}} -\E{S^{\alpha}}}{\varrho+\varsigma} A(x) B(x) (1+o(1)).
\EQNY
\ET
{\exxa (Third-order Hall-class)} Let $X$ be a random variable with a df $F$ such that, for some $\alpha, b>0, \varrho<0$ and $c,d\neq0$
\BQN\label{def:ThirdHall}
\bar F(x)= b x^{-\alpha} (1+c x^\varrho + d x^{2\varrho} (1+o(1))),\quad x\to\IF,
\EQN
i.e., $F$ is in the third-order Hall-class; see e.g., \cite{CaeiroG2008, CaideHZ2013, LiPX2011}. 
It follows then by Proposition \ref{Pro_inv} that $\bar F\in 3\RV_{-\alpha, \varrho, \varrho}$
with auxiliary functions $A$ and $B$ given by
\BQNY
A(x) =\frac{\varrho c x^\varrho}{1+cx^\rho}, \quad B(x) = \frac{2d}cx^\varrho.
\EQNY
An immediate application of \netheo{T5} with an independent scaling factor $S\in(0,1)$ yields that 
\BQNY
\bar H(x):=\pk{SX>x}=\E{S^\alpha}  b x^{-\alpha} \left(1+\frac{\E{S^{\alpha-\varrho}}}{\E{S^\alpha}}c x^\varrho + \frac{\E{S^{\alpha-2\varrho}}}{\E{S^\alpha}}d x^{2\varrho} (1+o(1))\right),\quad x\to\IF,
\EQNY
which together with Proposition \ref{Pro_inv} yields that the Value-at-Risk of $SX$ at level $q$, denoted by  $\VaR_q(SX)(:=\bar H^{\leftarrow}( 1-q))$, equals  with $c_q=\fracl {b\E{S^\alpha}}{1-q}^{1/\alpha}$
\BQNY
\frac{{\rm VaR}_q(SX)}{c_q}=1+\frac{c\E{S^{\alpha-\varrho}}}{\alpha\E{S^\alpha}} c_q^\varrho+\left(\frac12\fracl{c\E{S^{\alpha-\varrho}}}{\alpha\E{S^\alpha}}^2(1-\alpha+2\varrho)+ \frac{d\E{S^{\alpha-2\varrho}}}{\alpha\E{S^\alpha}} \right) c_q^{2\varrho}(1+o(1)),\quad q\uparrow1.
\EQNY
%
\BT\label{T6} Under the conditions as in \netheo{T2} (ii), we have 
\BQNY
\lefteqn{\frac{\pk{SX>x}}{\bar F(x)\bar G\left(1-1/{\varphi_t}\right)} = L_{\alpha } +M_{\alpha,1}  A (t) +N_{\alpha , 0, \varrho }\tilde A(\varphi_t)- \frac{\alpha L_{\alpha +1}}{\varphi_t}}\notag\\
&& +A (t) \Big(M_{\alpha,2}  A (t)+ Q_{\alpha}  B (t) + N_{\alpha,1, \varrho } \tilde A(\varphi_t)\Big)+N_{\alpha , 0, \varrho +\varsigma }\tilde A(\varphi_t) \tilde B(\varphi_t)(1+o(1)) \notag \\
 && -\frac\alpha{\varphi_t}\left(M_{\alpha+1,1} A (t)-  (\alpha +1) \frac{L_{\alpha +2}}{2\varphi_t}+ \Big(\frac{N_{\alpha +1,0, \varrho } }\alpha- L_{\alpha -\varrho +1}  \Big) \tilde A(\varphi_t)\right) (1+o(1))
\EQNY
with $\varphi_t=U(t)/a(t), t=1/\bar F(x)$ and $L_\alpha, M_{\alpha, l}, N_{\alpha, l,\varrho}, Q_{\alpha}$ given by \eqref{def: notaion}.
\ET
{\remark \label{remT6} We see that \netheo{T6} refines the second-order asymptotic expansions of deflated risks in \cite{HashorvaLP14} in a unified form, see Theorems 2.3 and 2.6 therein. }

\subsection{Asymptotic expansions of \HG risk measure}\label{Appl.HG}
%
\HG (H-G) risk measure is based on premium calculation principle via Orlicz norm which was first introduced by \cite{HaezendonckG1982}.  It is shown by \cite{BelliniR2008, BelliniRosazza2012} that H-G risk  measure is a law-invariant and coherent risk measure and thus an challenging alternative to Value-at-Risk and Expected Shortfall. \\
In what follows, with the aid of  \nekorr{Cor1} we shall establish the higher-order expansions of H-G risk measure which refine those by  \cite{MaoH2012, TangY2012}. It is shown (see Proposition 1.1 in \cite{MaoH2012} or \cite{TangY2012}) that, for a Young function  $\phi(t)=t^\kappa, \kappa\ge1$, and  $X$   a risk variable with
$\E{X_+^\kappa }<\IF$  and $\pk{X=x_F}=0$, then the H-G risk measure for $X$ at level $q\in(0,1)$, denoted by $H_q[X]$, is given by 
\BQN\label{def: HG}
H_q[X]=x+\fracl{\E{(X-x)_+^\kappa }}{1-q}^{1/\kappa},
\EQN
where $x=x(q)\in(-\IF, x_F)$ is the unique solution to the equation
\BQN\label{Eq: kq}
\frac{(\E{(X-x)_+^{\kappa-1}})^\kappa }{(\E{(X-x)_+^\kappa })^{\kappa-1}} = 1-q,\quad \kappa>1
\EQN
and $x= \VaR_q(X) (:= F^\leftarrow(q))$ for $\kappa=1$.  
\\ 
For simplicity of notation, denote, with $\xi_{\kappa ,\rho}, \widetilde M_{\kappa,1}, \Delta_\kappa $ given in \eqref{def: xi}
\BQN\label{def:c0123}
\left\{
\begin{array} {l} \c= \kappa\fracl{1-\kappa\gamma}{\kappa|\gamma|}^\kappa \xi_{\kappa ,0};\quad c_0= \frac{\ \c^\gamma}{1-\kappa\gamma}; \quad c_1=\frac1\rho\left(\ \c^\rho\frac{\xi_{\kappa ,\rho}}{\xi_{\kappa ,0}}-1\right)\\
c_2= \gamma  \c^{2\rho} \left(\frac1{2\gamma^2\rho^2}\left[(1-\gamma-2\rho)\frac{\xi_{\kappa ,2\rho}}{\xi_{\kappa ,0}}-2(1-\gamma-\rho)\frac{\xi_{\kappa ,\rho}}{\xi_{\kappa ,0}}+1-\gamma\right] \right. \\
\qquad+\Delta_\kappa \left[\kappa(\gamma+\rho)\widetilde M_{\kappa,1} +\left(\rho+\frac{\gamma-1}2\right)\Delta_\kappa +\frac1\gamma\right] \\
\left.\qquad+\kappa\widetilde M_{\kappa,1}\left[\frac {\kappa-1}2 \widetilde M_{\kappa,1} - \frac1{\gamma\rho}\left(\frac{\xi_{\kappa -1,\rho}}{\xi_{\kappa -1,0}}-1\right)\right]\right) + \c^\rho\frac{ \c^\rho-1}{\rho^2}\left(\frac{\xi_{\kappa ,\rho}}{\xi_{\kappa ,0}}-1\right) \\
c_3=\frac1{\rho+\eta}\left( \c^{\rho+\eta}\frac{\xi_{\kappa ,\rho+\eta}}{\xi_{\kappa ,0}}-1\right)
+\frac{ \c^{ \cL{\rho}}( \c^\eta-1)}{\rho\eta}\left(\frac{\xi_{\kappa ,\rho}}{\xi_{\kappa ,0}}-1\right).
\end{array}
\right.
\EQN
\BT\label{T3}
Let $H_q[X]$ be defined by \eqref{def: HG}. We have (i)  If $U\in3\RV_{\gamma, \rho, \eta}, 0<\kappa\gamma<1$ and $\rho, \eta\le0$ with auxiliary functions $A$ and $B$, then 
\BQNY
H_q[X] = c_0F^\leftarrow(q) \left(1+c_1\epsilon_q+ c_2\epsilon_q^2 (1+o(1)) +c_3 \epsilon_q \psi_q (1+o(1)) \right), \quad q\uparrow1.
\EQNY
(ii) 
If $x_F-U\in 3\RV_{\gamma, \rho, \eta}, \gamma<0$ and $\rho,\eta\le0$ with auxiliary functions $A$ and $B$, then
\BQNY
x_F-H_q[X] = c_0(x_F-F^\leftarrow(q))  \left(1+c_1\epsilon_q + c_2\epsilon_q^2 (1+o(1)) +c_3 \epsilon_q \psi_q (1+o(1)) \right), \quad q\uparrow1,
\EQNY
where  $\c, c_i, 0\le i\le 3$ are given as in \eqref{def:c0123} and
$\epsilon_q:=A(1/(1-q)), \psi_q:=B(1/(1-q))$.
\ET
{\remark\label{remHG} 
(i) Clearly, our \netheo{T3} provides the third-order asymptotics of $H_q[X]$ which is based on the bias-adjustment of the second-order ones; see Example \ref{eg_Burr} and \ref{eq_student}. 
\\
(ii) \netheo{T3} refines those in \cite{MaoH2012}. 
}

\subsection{Asymptotic expansions of expectile risk measure}\label{Appl.Expectiles}
Expectiles are first
introduced in \cite{NeweyP1987} in a statistical context. Namely, for a random variable $X$ 
with finite expectation 
the expectile $e_q=e_q[X], q\in [0,1]
$ is defined as the unique minimizer of an asymmetric quadratic loss
function as follows
\BQNY e_q
= \arg \min_{x\in\R}
\big(q\E{(X-x)_+^2} +(1-q)\E{(x-X)_+^2}\big)
\EQNY
or equivalently as the unique solution of the first-order condition
\BQN\label{def: expectile}
e_q   = \E{X}+ \frac{2q-1}{1-q}\E{(X-e_q)_+}, 
\EQN
see, e.g., \cite{BelliniB2014, BelliniKMR2014} for further discussions. 
Several generalizations of expectiles and its application are extensively studied in both statistical and actuarial literature; see, e.g., \cite{Gneiting2011, WangZ2014, Ziegel2014}. 

In the following, we investigate the higher-order expansions of $e_q$ extending those by \cite{BelliniB2014, MaoHN2015}. For simplicity of notation,  we denote for $0<\gamma<1, \rho, \eta\le 0$ and $D=({\gamma^{-1}}-1)^{-\rho}/(1-\gamma- \rho)$
 \BQNY
d_0 &=& \gamma({\gamma^{-1}}-1)^{\gamma}\E{X},\quad d_1=-2\gamma, \quad d_2=D+\frac{({\gamma^{-1}}-1)^{-\rho}-1}{\rho},\\
d_3 &=& 2(\gamma^2-\gamma),\quad d_4= D^2\left(\frac{\gamma-1}{2\gamma}+\frac\rho\gamma\right) + D\left(\left(\frac1\rho+ \frac1\gamma\right)\fracl\gamma{1-\gamma}^\rho -\frac1\rho\right)  ,\\
 d_5&=& -2(1+\rho)D - 2\frac{({\gamma^{-1}}-1)^{-\rho}-1}{{\gamma^{-1}}\rho}- 2({\gamma^{-1}}-1)^{-\rho},\\
 d_6&=&D\left(\frac{1-\gamma-\rho)({\gamma^{-1}}-1)^{-\eta}}{1-\gamma-\rho-\eta}+\frac{({\gamma^{-1}}-1)^{-\eta}-1}\eta\right) + \frac{({\gamma^{-1}}-1)^{-\rho-\eta}-1}{\rho+\eta},\\
d_7&=&
\frac{(\gamma^{-1}-1)^{2\gamma+1}}{2}(\gamma\E{X})^2,\quad d_8=-2\gamma^2(\gamma^{-1}-1)^{\gamma+1}\E{X},\\
d_9&=& \frac{2\gamma({\gamma^{-1}}-1)^{\gamma+1-\rho}}{1-\gamma-
\rho}\E{X}.
\EQNY

\BT\label{T4} If $U\in2\RV_{\gamma, \rho}, \gamma\in(0,1),
\rho\le0$ with auxiliary function $A$, then as $q\uparrow1$ \BQNY
e_q = \fracl\gamma{1-\gamma}^\gamma F^\leftarrow(q) \left(1+
\left(\frac{d_0}{F^\leftarrow(q)}  + d_1(1-q)  +d_2
\epsilon_q\right)(1+o(1))\right). \EQNY
If further
$U\in3\RV_{\gamma, \rho, \eta}, \ \gamma\in(0,1), \rho, \eta\le0$
with auxiliary functions $A$ and $B$. Then, as
$q\uparrow1$
\BQNY
\lefteqn{e_q= \fracl\gamma{1-\gamma}^\gamma F^\leftarrow(q) \left(1 +\frac{d_0}{F^\leftarrow(q)} + d_1(1-q) + d_2\epsilon_q \right.} \\
&& +  d_3 (1-q)^2 (1+o(1))+ \epsilon_q(d_4\epsilon_q + d_5 (1-q)  + d_6\psi_q)(1+o(1)) \\
&&+\left. \frac1{F^\leftarrow(q)}\left( \frac{d_7}{F^\leftarrow(q)} +
d_8(1-q)  +
d_9 \epsilon_q\right) (1+o(1))\right),
\EQNY
with
$\epsilon_q:=A(1/(1-q)), \psi_q:=B(1/(1-q))$
\ET
{\remark\label{remarkT4}  
(i) The speed of the second-order expansion is determined by $\max(\rho-1, 2\rho, \rho+\eta)$ for $\E{X}=0$, see Example \ref{eq_student}; otherwise by $\max(-2\gamma,2\rho, \rho+\eta)$ and thus we may neglect the terms related to $d_3, d_5$
and $d_8$, see Example  \ref{eg_Burr}. \\
(ii) \netheo{T4} extends Theorem ? in \cite{MaoHN2015}. Moreover, it is worth mentioning that their results for $\E{X}\neq0$ are only available for $\rho\ge -1$ since the location transformation will change the second-order parameter; see \cite{CaeiroG2008}}.

Next, we consider the case that $X\sim F$ which is in the Weibull MDA.
\BT\label{T7}
If $x_F-U\in 2\RV_{\gamma, \rho}, \gamma, \rho<0$ with auxiliary function $A$ such that, with a constant $C>0$
\BQNY
x_F-U(t) = Ct^\gamma\left(1+\frac{A(t)}\rho (1+o(1))\right),
\EQNY
then, with $\alpha=-1/\gamma, \, x_0=(\alpha+1)(x_F-\E X)$
\BQNY
x_F - e_q &=&(C^\alpha x_0(1-q))^{1/(\alpha+1)}\left(1-\frac{(C^\alpha x_0(1-q))^{1/(\alpha+1)}}{\alpha(x_F-\E{X})}(1+o(1)) \right.\\
&&\left.+\frac{(\alpha+1)(C/x_0)^{\alpha\rho/(\alpha+1)}}{\rho(\alpha+1-\alpha\rho)}A\left((1-q)^{-\frac\alpha{\alpha+1}}\right)(1+o(1))\right).
\EQNY
\ET
{\remark \label{rem3}
(i) \netheo{T7} generalizes Proposition 2.5 in \cite{BelliniB2014}. Moreover, one may refine the second-order results above by imposing $3RV$ conditions on $x_F-U$..\\
(ii) Numerous examples of $F$ that satisfy the conditions of \netheo{T6} and \netheo{T7} are presented in \cite{HashorvaLP14}. \\
(iii) One can follow the similar arguments for H-G and expectile risk measures above to consider the case that $X$ has a Weibull-type tails. }

\section{Examples and Numerical Analysis} \label{sec4}

{\exxa (Burr distribution)\label{eg_Burr}}  Let $X$ be a Burr distributed random variable with parameters $a, b>0$, i.e., 
$\bar F(x)=(1+x^a)^{-b}, x\ge0$. It follows that (cf. \cite{BarbeM2004})
\BQNY
\bar F(x)  =x^{-ab}\left(1-bx^{-a} +\frac{b(b+1)}{2}x^{-2a}(1+o(1))\right),\quad x\to\IF.
\EQNY
Consequently, it follows by Proposition \ref{Pro_inv} (
see also Table 1 in \cite{CaeiroG2008}) 
with $\alpha=-ab, \rho=-a$ that
\BQNY
U(t) = \overline F^\leftarrow(1/t) =t^{1/(ab)}\left(1-\frac{t^{-1/b}}a+\frac{1-a}{2a^2}t^{-2/b} (1+o(1))\right).
\EQNY
Consequently, $U\in3\RV_{1/(ab), -1/b, -1/b}$ with auxiliary functions $A, B$ given by
\BQNY
A(t)=\frac{t^{-1/b}}{ab-bt^{-1/b}},\quad B(t)=\frac{a-1}{a}t^{-1/b}.
\EQNY
For expectiles, note that  $\E{X}=a^{-1}B(b-1/a,1/a)$ for $ab>1$. By \netheo{T4} we have the first-, second- and third-order approximations of $e_q$ as $q\uparrow1$. In Figure \ref{fig1.1Burr} we take $a=2, b=1.5$ and we see the higher-order approximation of $e_q$ is more accurate than the lower-order ones.
\\
For H-G risk measure, we take $a=1/2, b=4$ and $\kappa=1.5$, and thus $\gamma=1/2, \rho=\eta=-1/4$. By \netheo{T3}
\BQNY
\frac{H_q[X]}{F^\leftarrow(q)}&=& 2.6935  \left( 1+1.9312\cdot\frac{(1-q)^{1/4}}{2(1-2(1-q)^{1/4})}+ \frac{(1-q)^{1/4}}2\left(-0.1822\cdot\frac{(1-q)^{1/4}}2 -1.2575(1-q)^{1/4}\right)(1+o(1))\right) \\
&=&2.6935  \left( 1+ 0.5116(1-q) +0.2587(1-q)^2 (1+o(1))\right), \quad q\uparrow1.
\EQNY
Therefore, we see that the first-, and second-order approximations  slightly underestimate the $H_q[X]$ while the third-order approximation has smaller error.

{\exxa (Student $t$-distribution)\label{eq_student}}  Let $X\sim
t_v$ be a student $t$-distributed random variable with $v>1$ degrees
of freedom. We have $\E{X}=0$ and its probability density function
$f$ is given by
\begin{align*} f(x) &= \frac{\Gamma((v+1)/2)}{\sqrt{v\pi} \Gamma(v/2)}
\left(1+ \frac{x^2}{v}\right)^{-(v+1)/2}\\
&=\frac{\Gamma((v+1)/2)}{\sqrt{v\pi} \Gamma(v/2)}v^{(v+1)/2} x^{-(v+1)}\left(1-\frac{v(v+1)}{2}x^{-2} + \frac{v^2(v+1)(v+3)}{8}x^{-4}(1+o(1))\right),\quad x\to\IF\\
&=: C_vx^{-k_1}(1+k_2x^{\rho\rq{}} +k_3x^{2\rho\rq{}}(1+o(1)))\quad {\rm with\ } C_v:=v^{v/2}/B(v/2,1/2).
 \end{align*}
It follows from \cL{\nelem{L1} and further by the dominated convergence theorem} that 
\BQNY
\overline F(x) &=& x f(x)\int_1^\IF\frac{f(tx)}{f(x)}\, dt \\
&=& x f(x)\int_1^\IF t^{-k_1}\left(1+ \frac{k_2 x^{\rho\rq{}}(t^{\rho\rq{}}-1)}{1+k_2x^{\rho\rq{}} +k_3x^{2\rho\rq{}}(1+o(1))} +k_3x^{2\rho\rq{}}(t^{2\rho\rq{}}-1)(1+o(1)) \right)\, dt \\
&=& \frac{xf(x)}{k_1-1} \left(1+\frac{\rho\rq{}}{k_1-1-\rho\rq{}}\frac{k_2x^{\rho\rq{}}}{1+k_2x^{\rho\rq{}}+k_3x^{2\rho\rq{}}} + \frac{\cL{2\rho\rq{}}}{k_1-1-2\rho\rq{}}k_3x^{2\rho\rq{}} (1+o(1))\right) \\
&=& \frac{C_v}v x^{-v} \Big( 1-\frac{v^2(v+1)}{2(v+2)} x^{-2} +\cL{\frac{v^3(v+1)(v+3)}{8(v+4)}}x^{-4}(1+o(1)) \Big),\quad x\to\IF.
\EQNY
Therefore, it follows from \neprop{Pro_inv} that (see also Table 1 in \cite{CaeiroG2008})
\BQNY
U(t)= \overline F^\leftarrow(1/t)= \fracl{C_vt}{v}^{1/v}\left(1-\frac{v(v+1)}{2(v+2)} \fracl{C_vt}{v}^{-2/v} -\frac{v^3(v+1)(v+3)}{8(v+2)^2(v+4)}
\fracl{C_vt}{v}^{-4/v} (1+o(1))\right).
\EQNY
Consequently, we have $U\in 3\RV_{1/v, -2/v, -2/v}$ with auxiliary functions
$A$ and $B$ given by
\BQNY
A(t)=\frac {(v+1)(C_vt/v)^{-2/v}}{v+2-v(v+1)(C_vt/v)^{-2/v}/2},\quad
B(t)=\cL{\frac{v^2(v+3)}{2(v+2)(v+4)}}\fracl{C_vt}{v}^{-2/v}.
\EQNY
We have by \netheo{T3} the second-order approximation with  a convergent rate  $\max(-1,-2/v)$. We obtain further by \netheo{T4} the third-order approximations of $e_q$ as $q\uparrow1$ with the speed of convergence of the second-order approximation as $\max(-2/v-1,-4/v)$. In Figure
\ref{fig1.2t}, we take $v=1.2$ and we see that the second- and third-order
approximations of $e_q$ is much better than the first-order approximations. Particularly, when $q=0.9979$,
the third-, second- and first-order evaluations are $(261.0483, 261.0483, 261.9426)$ for the true value $e_q=261.0483$. \\
For \HG risk measure, taking $v=2, \kappa=1.1$, and thus $C_2=1$. We have by \netheo{T3}
\BQNY
\frac{H_q[X]}{F^\leftarrow(q)}&=& 2.1044  \left( 1+0.6822\cdot\frac{3(1-q)}{4-3(1-q)}+ \frac{3(1-q)}4\left(-0.125\cdot \frac{3(1-q)}4+0.3998\cdot \frac56(1-q)\right)(1+o(1))\right) \\
&=&2.1044  \left( 1+ 0.5116(1-q) +0.5634(1-q)^2 (1+o(1))\right), \quad q\uparrow1.
\EQNY
Therefore, as in Example \ref{eg_Burr} we see that the first-, and second-order approximations  slightly underestimate the $H_q[X]$ while the third-order approximation has smaller error.

{\exxa (Beta distribution)\label{eg_beta}} Let $X\sim Beta(a,b), a, b>0$ with probability density function given by
\BQNY
f(x)=\frac1{B(a,b)} x^{a-1}(1-x)^{b-1}, \quad 0<x<1,\, a,\,b>0.
\EQNY
Note that
\BQNY
\bar F\left(1-\frac1t\right)=\int_{1-1/t}^1f(x)\,dx = \int_t^\IF \frac1{s^2}f\left(1-\frac1s\right)\,ds,
\EQNY
with
\BQNY
\frac1{t^2}f\left(1-\frac1t\right)=\frac{t^{-(b+1)}}{B(a,b)}
\left(1-\frac{a-1}t+\frac{(a-1)(a-2)}{2t^2}(1+o(1))\right),\quad t\to\IF.
\EQNY
Therefore, by the dominated convergence theorem
\BQNY
\bar F\left(1-\frac1t\right)=\frac{t^{-b}}{bB(a,b)}
\left(1-\frac{b(a-1)}{b+1}\frac1t+\frac{b(a-1)(a-2)}{2(b+2)}\frac1{t^2}(1+o(1))\right)=: g(t).
\EQNY
It follows by Proposition \ref{Pro_inv} and $1/(1-U(t))=g^\leftarrow(1/t)$ that
\BQNY
1-U(t) &=& \fracl t{bB(a,b)}^{-1/b}\left(1+\frac{a-1}{b+1} \fracl t{bB(a,b)}^{-1/b} \right. \\
&&+\left. \frac{a-1}{b+1} \left(\frac{a-1}{b+1}+\frac{a+b}{2(b+2)}\right)\fracl t{bB(a,b)}^{-2/b} (1+o(1))\right), \quad t\to\IF.
\EQNY
Consequently, we have $1-U\in 3\RV_{-1/b, -1/b,-1/b}$ with auxiliary functions $A$ and $B$ given by
\BQNY
A(t)&=& -\frac{a-1}{b(b+1)}\fracl t{bB(a,b)}^{-1/b}\Big/\left[1+\frac{a-1}{b+1}\fracl t{bB(a,b)}^{-1/b}\right] \\
B(t)&=& 2\left(\frac{a-1}{b+1}+\frac{a+b}{2(b+2)}\right)\fracl t{bB(a,b)}^{-1/b}.
\EQNY
For H-G risk measure, we take $a=3, b=6$ and $\kappa=2$, thus $\gamma=\rho=\eta=-1/6$. By \netheo{T3} 
\BQNY
\frac{1-H_q[X]}{1-F^\leftarrow(q)}&=&  0.8055\left(1- 0.9254 \cdot\frac1{21} \frac{((1-q)/28)^{1/6}}{1-2/7((1-q)/28)^{1/6}}\right)\\
&&\left. - \frac1{21}\fracl{1-q}{28}^{1/6} \left(\frac{1.98931}{21}\fracl{1-q}{28}^{1/6} + 1.1547\cdot \frac{95}{56}\fracl{1-q}{28}^{1/6}\right)(1+o(1))\right)\\
&=& 0.8055\left(1-0.0253(1-q)^{1/6} -0.0364(1-q)^{1/3}(1+o(1))\right), \quad q\uparrow1.
 \EQNY
Therefore, we see that the first-, and second-order approximations  slightly overestimate the $1-H_q[X]$ while the third-order approximation has smaller error.
In particular, for $a=b=1$,
we have $C=\alpha=1, \rho=-\IF$ \BQNY
1-e_q=\sqrt{1-q}\big(1-2\sqrt{1-q}(1+o(1))\big), \quad q\uparrow1
\EQNY which  coincides with the true value of $e_q =
(q-\sqrt{q-q^2})/(2q-1)$ (see Example 3.1 in \cite{BelliniB2014}). In
Figure \ref{fig1.5}, we take $(a, b)=(2, 3)$ and we see that our
second-order approximation of $e_q[X]$ performs very well.

{
\begin{figure}[htbp]
\begin{center}
\includegraphics[height=3.5cm, width=5cm]{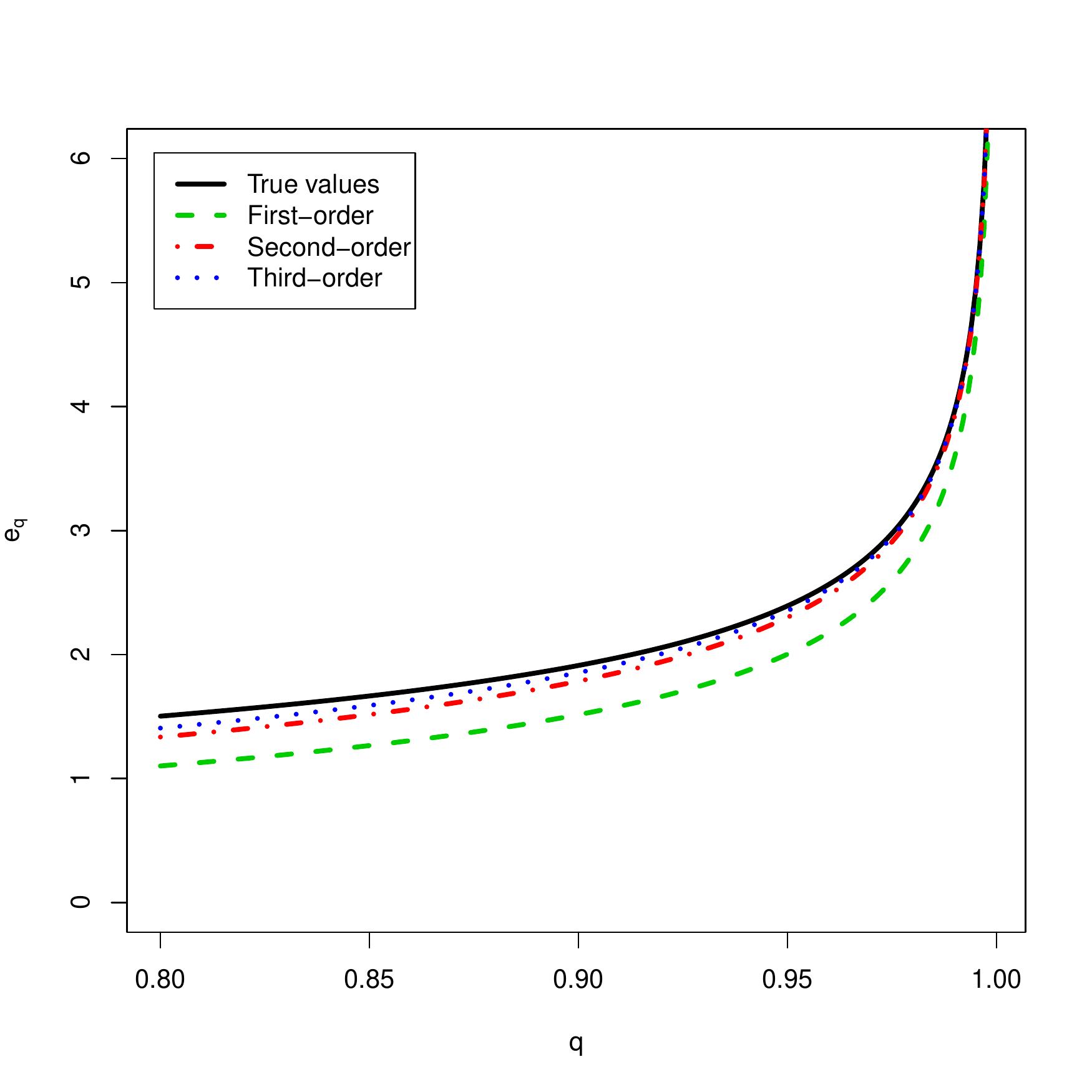}
\includegraphics[height=3.5cm, width=5cm]{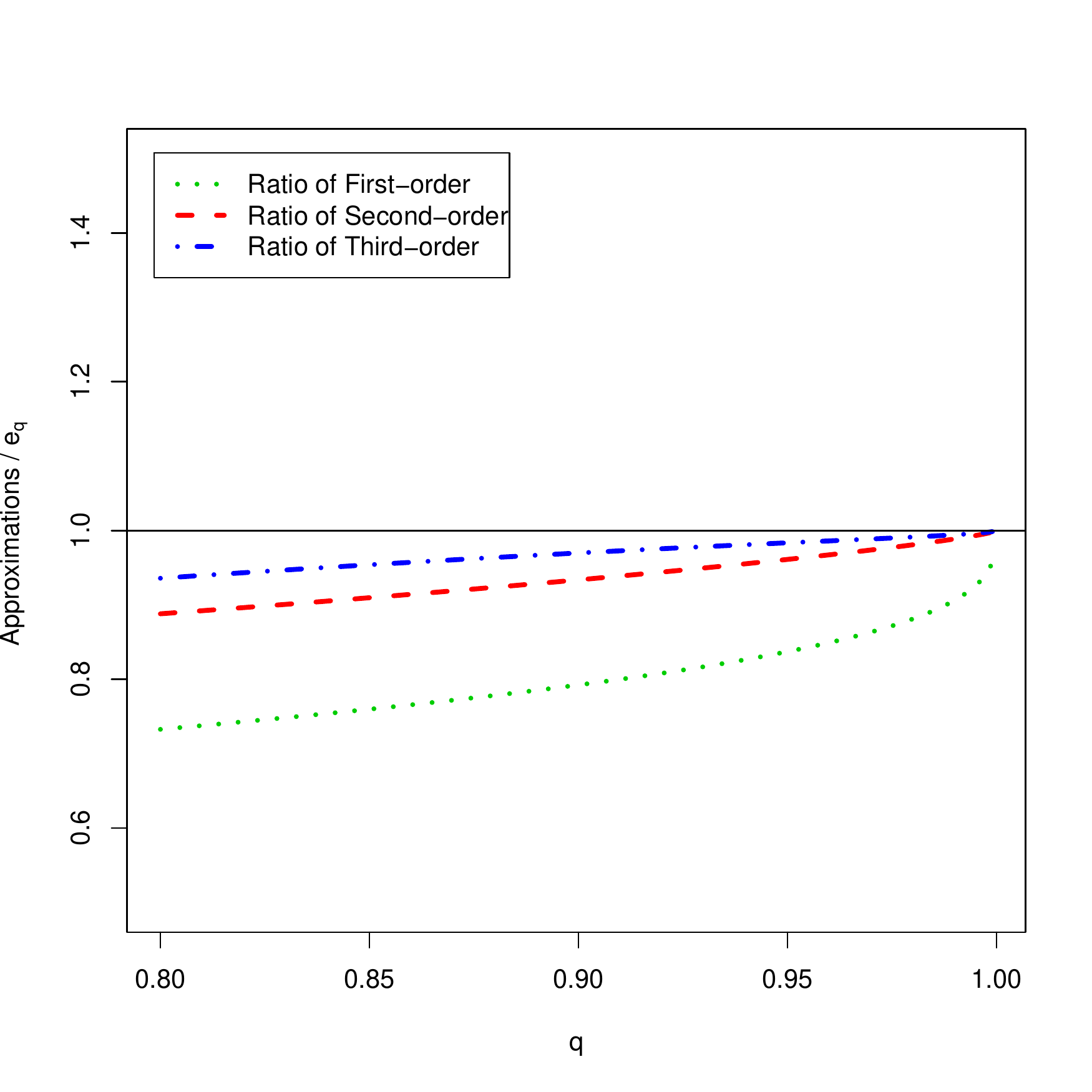}
\caption{Approximations of expectile $e_q[X], q\uparrow1$ (left) and
Ratios of approximations divided by the true values of $e_q[X]$
(right) for $X\sim Burr(a, b)$ with $(a, b)=(2, 1.5)$.
 } \label{fig1.1Burr}
\end{center}
\end{figure}

\begin{figure}[htbp]
\begin{center}
\includegraphics[height=3.5cm, width=5cm]{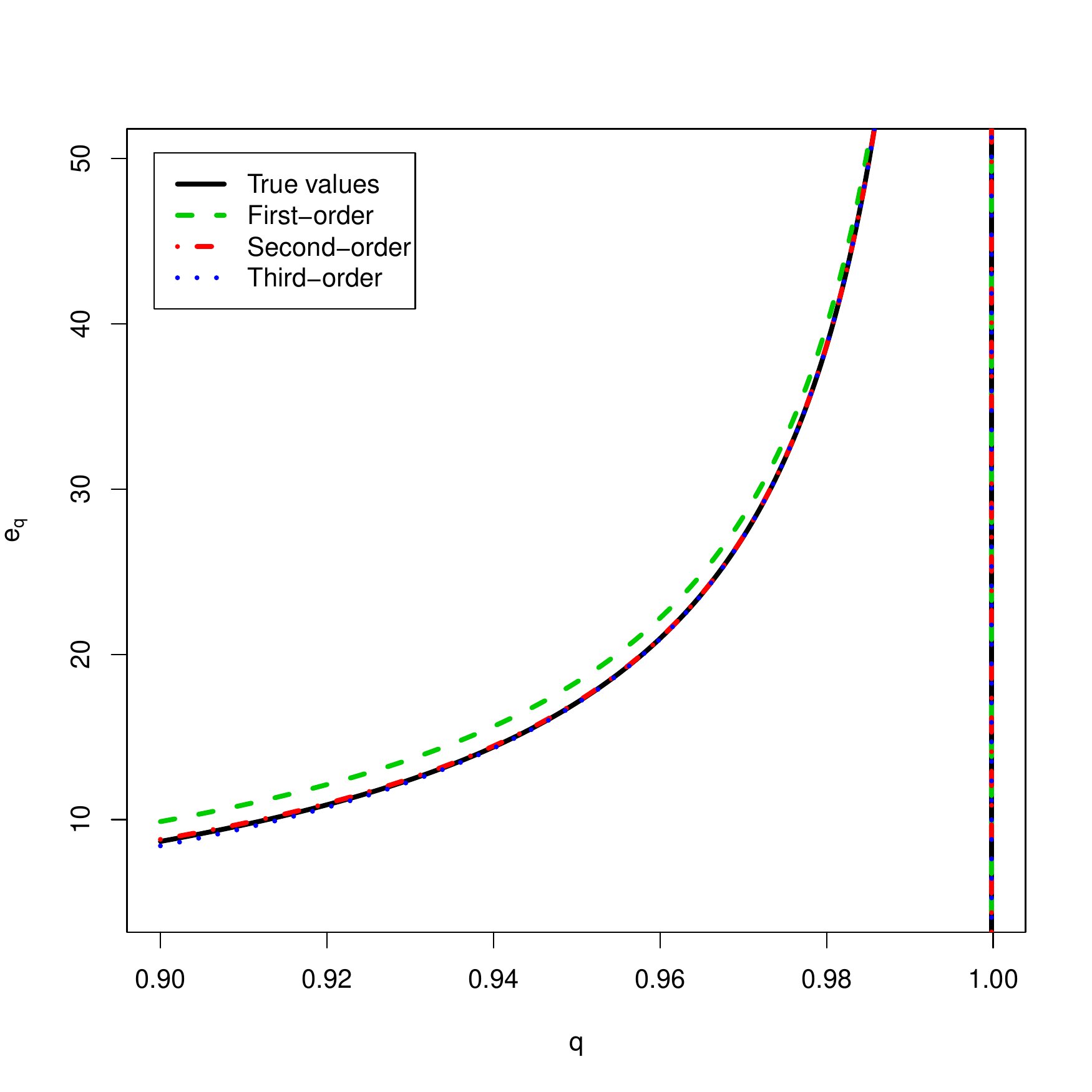}
\includegraphics[height=3.5cm, width=5cm]{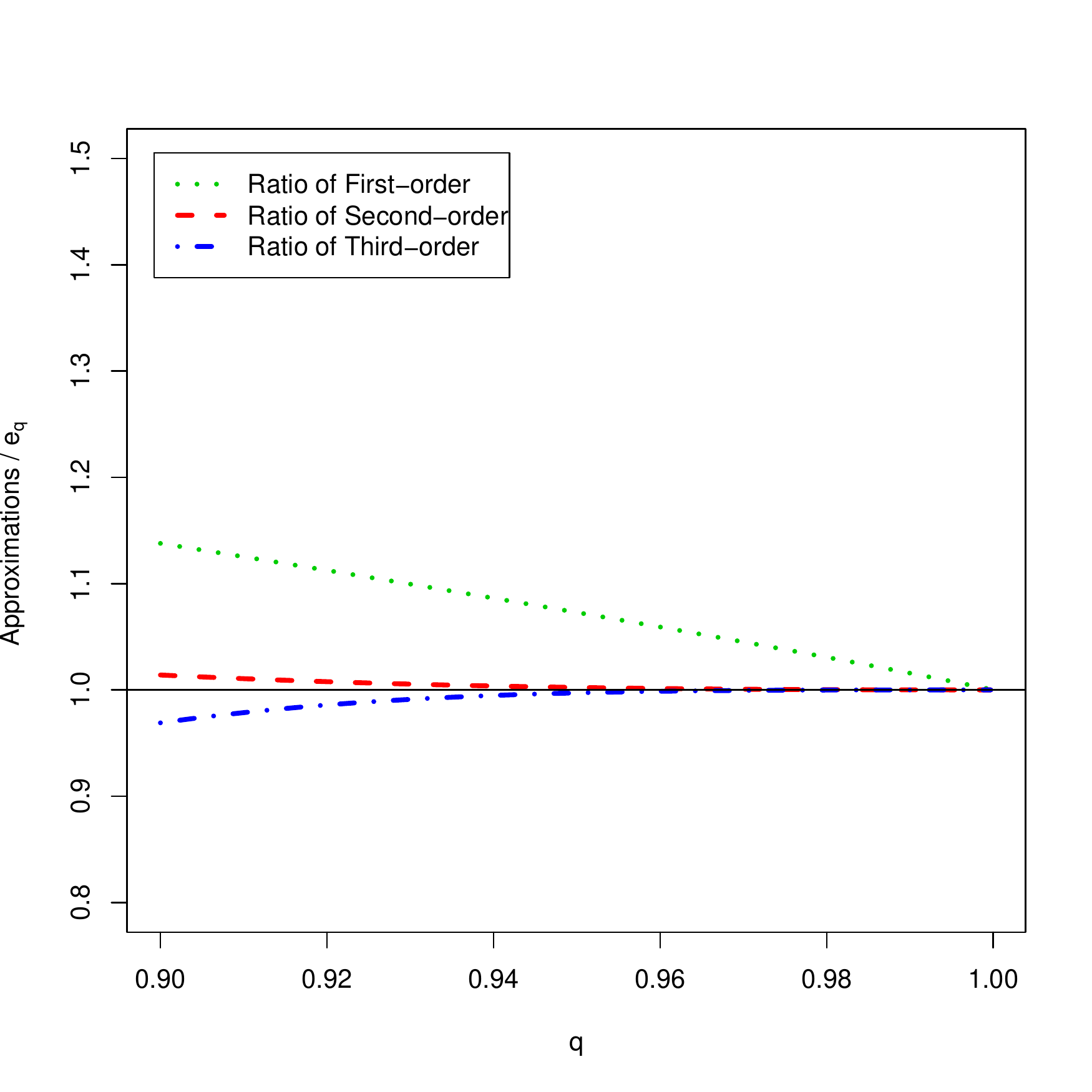}
\caption{Approximations of expectile $e_q[X], q\uparrow1$ (left) and
Ratios of approximations divided by the true values of $e_q[X]$
(right) for $X\sim t_v$ with $v=1.2$.
 } \label{fig1.2t}
\end{center}
\end{figure}

\begin{figure}[htbp]
\begin{center}
\includegraphics[height=3.5cm, width=5cm]{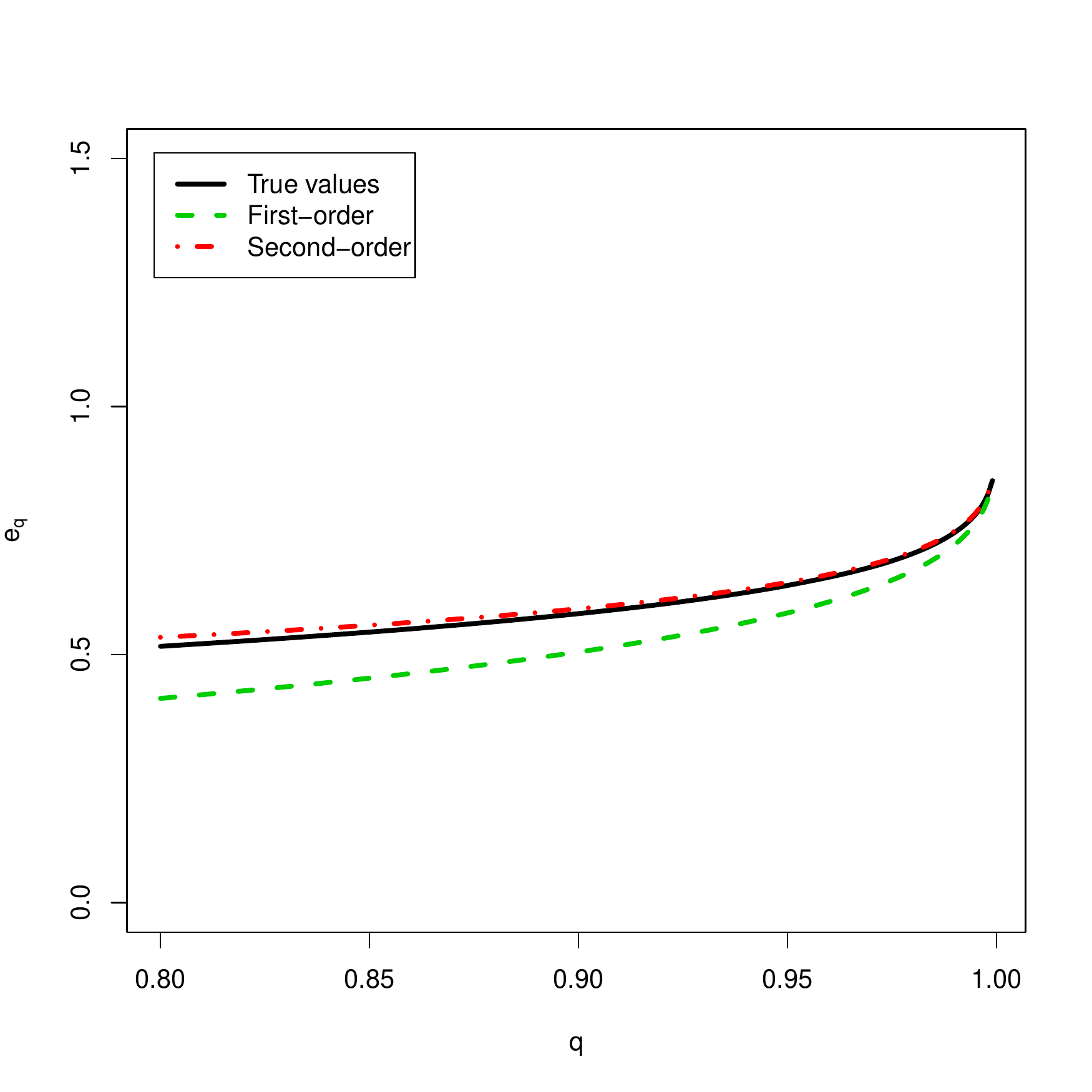}
\includegraphics[height=3.5cm, width=5cm]{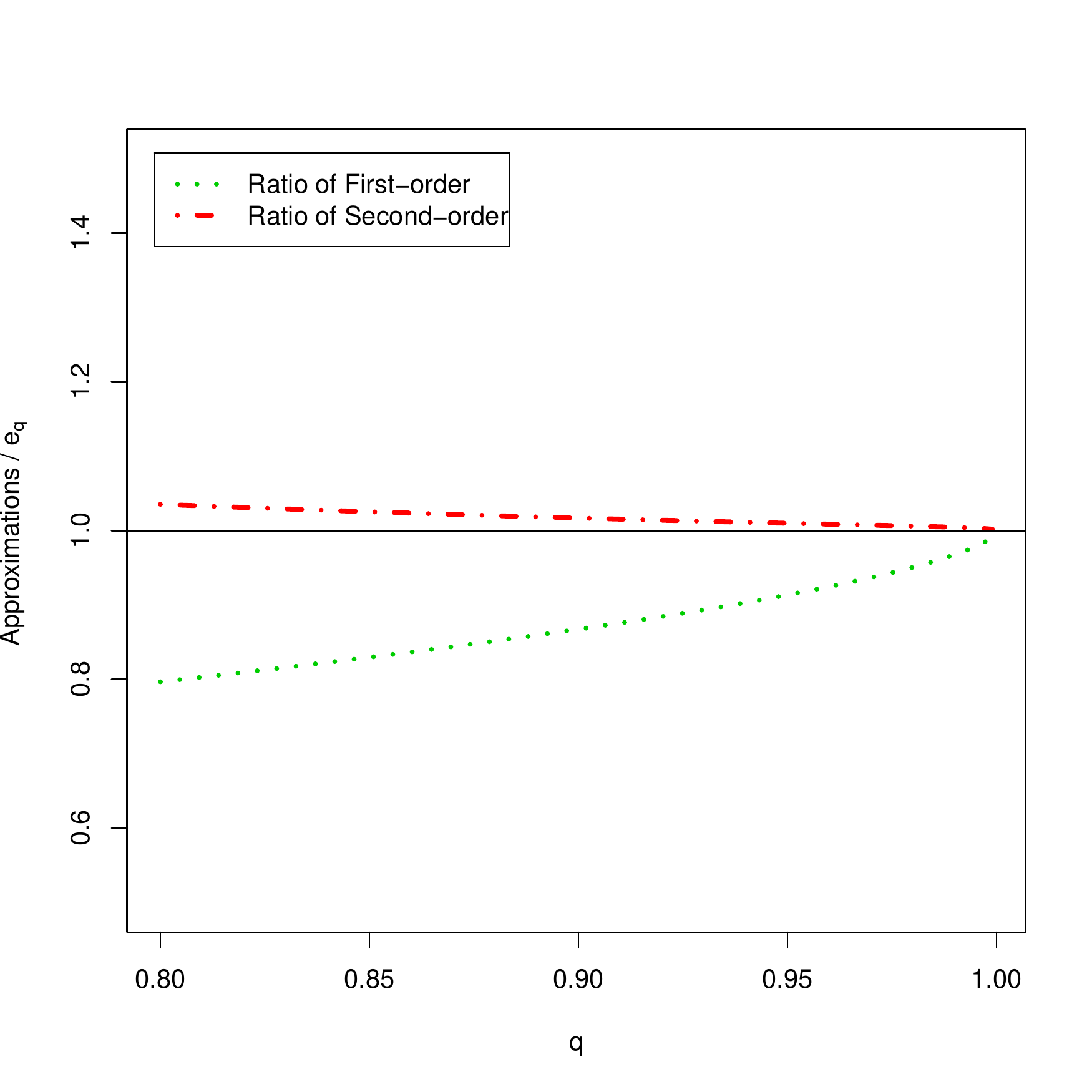}
\caption{Approximations of expectile $e_q[X], q\uparrow1$ (left) and
Ratios of approximations divided by the true values of $e_q[X]$
(right) with $X\sim Beta(2,3)$.
 } \label{fig1.5}
\end{center}
\end{figure}
}

\section{Proofs}\label{sec5}

\prooftheo{T1} Noting that $S$ and $X$ are independent, we have
\BQNY
\frac{\E{X^\kappa\I{SX>x}}}{\bar F(x)} &=& \frac1{\bar F(x)}\int_0^1 \E{X^\kappa\I{sX>x}}\, dG(s)\\
&=&\frac1{\bar F(x)}\int_0^1 \int_{x/s}^\IF y^\kappa\, dF(y)\, dG(s)\\
&=& x^\kappa\int_0^1  s^{-\kappa} \frac{\bar F(x/s)}{\bar F(x)}\left( 1+\kappa \int_{1}^\IF
y^{\kappa-1}\frac{\bar F((x/s)y)}{\bar F(x/s)}\,dy \right)\, dG(s).
\EQNY
It follows from Lemma 5.2 in \cite{DraismadPP1999}, for any given $\ve >0$, there exists $x_0>0$ such that for all $x>x_0$ and all $s\in(0, 1)$ and $y>1$
\BQN\label{Ineq.Second}
\left\{\begin{array}{l}
\Abs{\frac{\bar F(x/s)}{\bar F(x)}- s^{\alpha }\left(1+\frac{s^{-\varrho }-1}{\varrho }  A (x)\right)}\le \ve \abs{ A (x)} (1+s^{\alpha } +2s^{\alpha -\varrho -\ve}) \\
\Abs{\frac{\bar F((x/s)y)}{\bar F(x/s)} - y^{-\alpha }\left(1+\frac{y^{\varrho }-1}{\varrho }   A (x/s)\right) }\le \ve\abs{ A (x/s)}(1+y^{-\alpha } +2y^{-\alpha +\varrho +\ve}).
\end{array}
\right.
\EQN
Further a straightforward application of  Potter bounds  for $ A(\cdot) $  (cf. Proposition B.1.9 in \cite{deHaanF2006}) yields that $ A (x/s)/ A (x) \le (1+\ve) s^{-\varrho -\ve}$. Consequently, by the dominated convergence theorem
\BQNY
\lefteqn{\frac{\E{X^\kappa\I{SX>x}}}{x^\kappa\bar F(x)}= \int_0^1 s^{\alpha -\kappa}\left(1+\frac{s^{-\varrho }-1}{\varrho }  A (x)(1+o(1))\right)} \\
&&\times \left(1+\kappa \int_1^\IF y^{\kappa-\alpha -1}\left(1+\frac{y^{\varrho }-1}{\varrho } s^{-\varrho } A (x) (1+o(1))\right)\, dy  \right)\,dG(s)
\EQNY
where the $o(1)$-terms are uniform for all $s\in(0, 1)$ and $y\in(1,\IF)$. We complete the proof of \eqref{Expan.Second}. \\
To prove \eqref{Expan.Third}, 
we use  by \nelem{L1}
\BQNY
\lefteqn{\Abs{\frac{\bar F(x/s)}{\bar F(x)} - s^{\alpha }\left(1+\frac{s^{-\varrho }-1}{\varrho }  A (x)+ \frac{s^{-\varrho -\varsigma }-1}{\varrho +\varsigma }  A (x)  B (x)\right) }}\\
&&\le \ve \abs{ A (x) }\abs{  B (x)}(1+s^{\alpha } +2s^{\alpha -\varrho }+ 4s^{\alpha -\varrho -\varsigma -\ve})
\EQNY
\BQNY
\lefteqn{\Abs{\frac{\bar F((x/s)y)}{\bar F(x/s)} - y^{-\alpha }\left(1+\frac{y^{\varrho }-1}{\varrho } A (x/s) + \frac{y^{\varrho +\varsigma }-1}{\varrho +\varsigma } A (x/s)  B (x/s)\right)}} \\
&&\le \ve \abs{ A (x)}\abs{  B (x)}(1+y^{-\alpha } +2y^{-\alpha +\varrho } +4y^{-\alpha +\varrho +\varsigma +\ve})
\EQNY
and
\BQNY
\Abs{\frac{ A (x/s)}{ A (x)} -
s^{-\varrho }\left(1+\frac{s^{-\varsigma }-1}{\varsigma } B (x) \right) }&\le& \ve \abs{ B (x)}(1+s^{-\varrho} +2s^{-\varrho-\-\varsigma-\ve})\\
\frac{ A (x/s) B (x/s)}{ A (x) B (x)} &\le& (1+\ve)s^{-\varrho -\varsigma -\ve}.
\EQNY
Thus again using the dominated convergence theorem
\BQNY
\lefteqn{\frac{\E{X^\kappa\I{SX>x}}}{x^\kappa\bar F(x)}= \int_0^1 s^{\alpha -\kappa}\left(1+\frac{s^{-\varrho }-1}{\varrho }  A (x)+ \frac{s^{-\varrho -\varsigma }-1}{\varrho +\varsigma }  A (x)  B (x)(1+o(1))\right)} \\
&&\times \left(1+\kappa\int_1^\IF y^{\kappa-\alpha -1}\left(1+\frac{y^{\varrho }-1}{\varrho } A \fracl xs + \frac{y^{\varrho +\varsigma }-1}{\varrho +\varsigma } A \fracl xs  B \fracl xs (1+o(1))\right)\,dy\right)\,dG(s)\\
&=& \frac{\alpha }{\alpha -\kappa}\int_0^1 s^{\alpha -\kappa}\left(1+\frac{s^{-\varrho }-1}{\varrho }  A (x)+ \frac{s^{-\varrho -\varsigma }-1}{\varrho +\varsigma }  A (x)  B (x)(1+o(1))\right) \\
&&\times \left(1+\frac{\kappa s^{-\varrho }}{\alpha (\alpha -\kappa-\varrho )}  A (x) + \frac{\kappa}{\alpha }\left(\frac{s^{-\varrho }(s^{-\varsigma }-1)}{(\alpha -\kappa-\varrho )\varsigma } + \frac{s^{-\varrho -\varsigma }}{\alpha -\kappa-\varrho -\varsigma }\right)  A (x) B (x)(1+o(1))\right)\,dG(s)\\
&=: & \frac{\alpha }{\alpha -\kappa} \big(d_{0,\kappa} + d_{1,\kappa} A (x) + d_{2,\kappa}  A ^2(x)(1+o(1)) + d_{3,\kappa}  A (x) B (x)(1+o(1))\big)
\EQNY
establishing our proof with elementary consideration.
\QED

\prooftheo{T2} Letting $t=1/\bar F(x)$ and thus $x= U(t)$ for large $t$, we have
\BQN \label{MEF_decom}
\frac{\E{X^\kappa\I{SX>x}}}{x^\kappa\bar F(x)}&=&\int_0^1\fracl{U(t/s)}{U(t)}^\kappa \pk{S> 1- \frac{U(t/s)- U(t)}{U(t/s)}} \, ds \notag\\
&=&   \int_0^1\fracl{U(t/s)}{U(t)}^\kappa \pk{S> 1- \frac{U(t/s)- U(t)}{a(t)}\frac{a(t)}{U(t/s)}} \, ds\notag\\
&=&  \bar G\left(1-\frac1{\varphi_t}\right)\int_0^1 \left(1+\frac{\PF}{\varphi_t}\right)^\kappa \frac{\bar G\left(1-\frac{\PF}{\PF+\varphi_t}\right)}{\bar G\left(1-\frac1{\varphi_t}\right)}\,ds\notag\\
&=&  \bar G \left(1-\frac1{\varphi_t}\right)\int_0^1 ({\PF})^{\alpha }\left(1+\frac{\PF}{\varphi_t}\right)^{\kappa-\alpha }
\frac{L\left(1+\frac{\varphi_t}{\PF}\right)}{L\left(\varphi_t\right)}\,ds 
\EQN
with 
\BQNY
\PF=\frac{U(t/s)-U(t)}{a(t)}, \quad \varphi_t=\frac{U(t)}{a(t)},\quad L(z)= z^{\alpha} \bar G\left(1-\frac1z\right).
\EQNY
Further 
\BQN\label{decomp.1}
\lefteqn{\frac{\E{X^\kappa\I{SX>x}}}{x^\kappa \bar F(x) \bar G \left(1-1/{\varphi_t}\right)}= \int_0^1(D_\gamma(1/s))^{\alpha }\,ds + \int_0^1\left(({\PF})^{\alpha }-(D_\gamma(1/s))^{\alpha }\right)\,ds} \notag\\
&&\quad+\int_0^1({\PF})^{\alpha }\left(\left(1+\frac{\PF}{\varphi_t}\right)^{\kappa-\alpha }-1\right)\,ds + \int_0^1 ({\PF})^{\alpha }\left(1+\frac{\PF}{\varphi_t}\right)^{\kappa-\alpha }
\left(\frac{L\left(1+\frac{\varphi_t}{\PF}\right)}{L\left(\varphi_t\right)}-1\right)\,ds \notag\\
&&=: L_\alpha +\int_0^1(I_{1t}(s)+I_{2t}(s)+I_{3t}(s))\,ds.
\EQN
Next, we will present the proofs of \eqref{Expan.Second.GW} and \eqref{Expan.Third.GW} by dealing with $I_{it}(s), i=1,2,3$ one by one.\\
\underline{ (i) Proof of \eqref{Expan.Second.GW}}. It follows from Lemma 5.2 in \cite{DraismadPP1999} that, for any given $\ve\in(0,1)$, there exists $t_0=t_0(\ve) >0$ such that for all $t>t_0$ and all $s\in(0,1)$ we have  $\abs{\PF-D_\gamma(1/s)}\le \ve(1+s^{-\gamma}+2s^{-\gamma-\ve}) $. Furthermore, by Taylor's expansion $(1+x)^{\alpha }= 1+\alpha  x(1+o(1))$ for smaller $|x|$
\BQNY
\lefteqn{\Abs{\frac{I_{1t}(s)}{ A (t)}} = (D_\gamma(1/s))^{\alpha } \Abs{\frac{\left(1+\frac{\PF-D_\gamma(1/s)}{D_\gamma(1/s)}\right)^{\alpha } -1}{ A (t)}}}\\
&&\le \alpha  (D_\gamma(1/s))^{\alpha -1}\Abs{\frac{\PF-D_\gamma(1/s)}{ A (t)}} \big(1+\ve(1+s^{-\gamma}+2s^{-\gamma-\ve}) \big)\\
&&\le\alpha  (D_\gamma(1/s))^{\alpha -1} \left(H_{\gamma, \rho}(1/s) + \ve (1+s^{-\gamma} +2s^{-\gamma-\rho-\ve})\right)\big(1+\ve(1+s^{-\gamma}+2s^{-\gamma-\ve}) \big).
\EQNY
which is integrable in $(0,1)$. Thus by the dominated convergence theorem
\BQN\label{I1-second}
\int_0^1 I_{1t}(s)\,ds=  A (t) \alpha \int_0^1  (D_\gamma(1/s))^{\alpha -1} H_{\gamma, \rho}(1/s)\,ds (1+o(1)) =: M_{\alpha,1}A(t) (1+o(1)),\quad t\to\IF.
\EQN
For $I_{2t}(s)$, noting that $\varphi_t\to\IF$ as $t\to\IF$ and using for small $x\ge0$ that $\abs{(1+x)^l-1} \le 2\abs l x$
\BQN\label{Ineq.Power}
\varphi_t \abs{I_{2t}(s)} \le 2\abs{\kappa-\alpha } (q_t(s))^{\alpha +1}.
\EQN
Thus, using again the dominated convergence theorem, we have
\BQN\label{I2-second}
\int_0^1 I_{2t}(s)\,ds= \frac{1}{\varphi_t} (\kappa-\alpha)L_{\alpha +1}  (1+o(1)),\quad t\to\IF.
\EQN
It remains to deal with the third term $I_{3t}(s)$. By \eqref{Ineq.Power}, for large $t$ and all $s\in(0,1)$
\BQNY
\Abs{\frac{I_{3t}(s)}{\tilde A(\varphi_t)}} \le\frac{(\PF)^{\alpha }}{\abs{\tilde A(\varphi_t)}}\Abs{\frac{L\left(1+\frac{\varphi_t}{\PF}\right)}{L\left(\varphi_t\right)}-1} \left(1+2\abs{\kappa-\alpha } \frac{\PF}{\varphi_t}\right).
\EQNY
We consider the two cases a) $\gamma<0$ and b) $\gamma=0$ separately. \\
a) For $\gamma<0$, recalling that $1/\varphi_t+1/\PF>-\gamma$, using again Lemma 5.2 in \cite{DraismadPP1999} for $L\in 2\RV_{0,\varrho }$ that, for any given $\ve>0$, there exists $t_0=t_0(\ve) >0$ such that for all $\varphi(t)>t_0$ and all $s\in(0,1)$
\BQNY
\lefteqn{\Abs{\frac{I_{3t}(s)}{\tilde A(\varphi_t)}} \le (\PF)^{\alpha } \left(1+2\abs{\kappa-\alpha } \frac{\PF}{\varphi_t}\right)}\\
&&\times \left( \frac{(1/\varphi_t+1/\PF)^{\varrho } -1}{\varrho }+ \ve\left(1+\left(\frac1{\varphi_t}+\frac1\PF\right)^{\varrho }\expon{\ve\Abs{\ln \left(\frac1{\varphi_t}+\frac1\PF\right)}}\right) \right)
\EQNY
which is integrable. \\
b) For $\gamma=0$, it follows from Lemma 5.2 in \cite{DraismadPP1999} that  $|I_{3t}(s)/\tilde A(\varphi_t)|$ is integrable over $\mathbb S_t=\{s\in(0,1): \min(\varphi_t, 1+\varphi_t/\PF)>t_0\}$. And on $\mathbb S_t^c=\{s\in(0,1): \varphi_t>t_0>1+\varphi_t/\PF\}$, we use the similar arguments for proving Theorem 2.3 in \cite{HashorvaLP14} (see (5.6) therein) as follows. \\
Clearly $L(1+\varphi_t/\PF)\le (1+\varphi_t/\PF)^{\alpha } \le t_0^{\alpha }$. Moreover, by Theorem 1.10 in \cite{Gelukd1987} that $\lim_{t\to\IF}(L(t) -L(\IF))/\tilde A(t)=1/\varrho $ where  $L(\IF):=\lim_{t\to\IF} L(t)$ and $\varrho <0$ implying $L(\IF)>0$. Therefore $L(\varphi_t)>L(\IF)/2$ for large $t$. Meanwhile, by Potter bounds, for any given $\ve\in(0, 1)$ and $\varphi_t>t_0>1+\varphi_t/\PF$
\BQN\label{ineq.A}
|\tilde A(\varphi_t)| \ge |\tilde A((t_0-1) \PF)| \ge (1-\ve)( \PF)^{\varrho -\ve}|\tilde A(t_0-1)|.
\EQN
Consequently
\BQNY
\Abs{\frac{I_{3t}(s)}{\tilde A(\varphi_t)}} \le (\PF)^{\alpha -\varrho+\ve} \left(1+2\abs{\kappa-\alpha } \frac{\PF}{t_0}\right) \frac{2t_0^{\alpha }/L(\IF)+1}{(1-\ve)|\tilde A(t_0-1)|}
\EQNY
holds on $\mathbb S_t^c$ whose Lebesgue measure goes to zero as $t\to\IF$. \\
Hence, a straightforward application of the dominated convergence theorem to $I_{3t}(s)$ for both a) $\gamma<0$ and b) $\gamma=0$ yields that
\BQN\label{I3-second}
\frac{1}{\tilde A(\varphi_t)} \int_0^1I_{3t}(s)\,ds &=&  \int_0^1 ({\PF})^{\alpha }\left(1+\frac{\PF}{\varphi_t}\right)^{\kappa-\alpha }
\frac1{\tilde A(\varphi_t)}\left(\frac{L\left(1+\frac{\varphi_t}{\PF}\right)}{L\left(\varphi_t\right)}-1\right)\,ds
\notag\\
&=& \int_0^1\left(D_\gamma\fracl1s\right)^{\alpha }\frac{(D_\gamma(1/s))^{-\varrho }-1}{\varrho }\,ds (1+o(1))\notag\\
&=:& N_{\alpha,0,\varrho}(1+o(1)),\quad t\to\IF
\EQN
which together with \eqref{decomp.1}, \eqref{I1-second} and \eqref{I2-second} establishes the proof of \eqref{Expan.Second.GW}. \\
\underline{(ii) Proof of \eqref{Expan.Third.GW}}. In the following, we use the same notation aforementioned as  before. 
For $I_{1t}(s)$ given in \eqref{decomp.1}, using  $(1+x)^{\alpha } = 1+\alpha  x+ c_{\alpha,2}x^2(1+o(1))$ for smaller $x$ and \nelem{L1}, \cL{there exists a constant $C>0$,} for any $\ve>0$ there exists some $t_0=t_0(\ve)>1$ such that for all $t>t_0$ and  $s\in(0,1)$
\BQNY
\lefteqn{\frac{\Abs{I_{1t}(s)/ A (t)-\alpha  (D_\gamma(1/s))^{\alpha -1} H_{\gamma, \rho}(1/s)}}{\abs{ A (t)}+\abs{ B (t)}}}\\
 &&\le\alpha  (D_\gamma(1/s))^{\alpha -1} \frac{\Abs{\frac{\PF-D_\gamma(1/s)}{ A (t)}-H_{\gamma, \rho}(1/s)}}{\abs{ B (t)}}\frac{\abs{ B (t)}}{\abs{ A (t)}+\abs{ B (t)}}\\
 &&\quad + c_{\alpha,2}(D_\gamma(1/s))^{\alpha -2} \fracl{\PF-D_\gamma(1/s)}{ A (t)}^2 \frac{\abs{ A (t)}}{\abs{ A (t)}+\abs{ B (t)}} \big(1+\ve(1+s^{-\gamma}+2s^{-\gamma-\ve})\big)\\
 &&\le \alpha  (D_\gamma(1/s))^{\alpha -1} \big(R_{\gamma, \rho, \eta}(1/s) + \ve(1+s^{-\gamma} +2s^{-\gamma-\rho} + 4s^{-\gamma-\rho-\eta-\ve} \cL{+s^{-C}\I{\gamma=\rho=0}})\big) \\
 &&\quad +  c_{\alpha,2}(D_\gamma(1/s))^{\alpha -2}\big(H_{\gamma, \rho}(1/s)+\ve(1+s^{-\gamma} +2s^{-\gamma-\rho-\ve})\big)^2 \big(1+\ve(1+s^{-\gamma}+2s^{-\gamma-\ve})\big).
\EQNY
Consequently, again by the dominated convergence theorem
\BQN\label{I1-third}
\int_0^1 I_{1t}(s)\,ds=  A (t) \Big(M_{\alpha, 1}+  B (t)Q_\alpha (1+o(1)) + A (t)M_{\alpha,2} (1+o(1))\Big) ,\quad t\to\IF.
\EQN
For $I_{2t}(s)$, note that $\abs{(1+x)^p-1-px}\le Cx^2$ for small
 $x\ge0$ and $C=C_p>0, p\in\R$. It follows that, for any $\ve>0$ there exists $t_0=t_0(\ve)>0$ such that for $t>t_0$ and $s\in(0,1)$
\BQNY
\lefteqn{ \varphi_t\Abs{\varphi_t I_{2t}(s) - (\kappa-\alpha)(q_t(s))^{\alpha +1} }
} \\
&& = \varphi_t^2(q_t(s))^{\alpha }\Abs{\left(1+\frac{q_t(s)}{\varphi_t}\right)^{\kappa-\alpha } -1-(\kappa-\alpha )\frac{q_t(s)}{\varphi_t} } \\
&&\le C(q_t(s))^{\alpha +2}.
\EQNY
Consequently, again by the dominated convergence theorem
\BQN\label{I2-third}
\lefteqn{ \int_0^1 I_{2t}(s)\,ds =\frac{\kappa-\alpha}{\varphi_t} \int_0^1(q_t(s))^{\alpha +1} \left( 1 + \frac{\kappa-\alpha -1}{2}\frac{q_t(s)}{\varphi_t}(1+o(1))\right)\,ds}\notag\\
&&=\frac{\kappa-\alpha}{\varphi_t} \int_0^1\left(D_\gamma\fracl1s + H_{\gamma, \rho}\fracl1s  A (t) (1+o(1))\right)^{\alpha +1} \left(1+\frac{\kappa-\alpha -1}{2}\frac{q_t(s)}{\varphi_t}(1+o(1))\right)\,ds \notag\\
&&=\frac{\kappa-\alpha}{\varphi_t}\left(L_{\alpha +1}+  A
(t)M_{\alpha+1,1} (1+o(1))+ \frac{\kappa-\alpha
-1}{2\varphi_t}L_{\alpha +2} (1+o(1)) \right), \EQN where the
$o(1)$'s terms in the integral are uniform for $s\in(0,1)$, and the
following are the same unless otherwise stated. \\
 Next, we deal
with $I_{3t}(s)$. 
First note that $\abs{\tilde B}\in \RV_\varsigma$. Similar arguments for \eqref{ineq.A} yield that, 
for any given $\ve\in(0,1)$, there exists some $t_0=t_0(\ve)>0$ such
that for $\varphi(t)>t_0$ 
\BQNY |\tilde B(\varphi_t)| \ge |\tilde
B((t_0-1)q_t(s))| \ge (1-\ve)(q_t(s))^{\varsigma -\ve}|\tilde
B(t_0-1)|. 
\EQNY
Using further \nelem{L1} to analyze  the two cases: 
a) $\gamma<0$ and b) $\gamma=0$ above,  we see that \BQNY \lefteqn{ \frac1{\abs{\tilde
B(\varphi_t)}} \Abs{\frac{I_{3t}(s)}{\tilde A(\varphi_t)} -
(q_t(s))^{\alpha
}\left(1+\frac{q_t(s)}{\varphi_t}\right)^{\kappa-\alpha }
\frac{(1/\varphi_t+1/q_t(s))^{\varrho }-1}{\varrho}}
} \\
&& \le (\PF)^{\alpha } \left(1+2\abs{\kappa-\alpha } \frac{\PF}{\varphi_t}\right)\\
&&\times \left[\left(\frac{(1/\varphi_t+1/\PF)^{\varrho} -1}{\varrho}+ \ve\left(1+(q_t(s) )^{-\varrho } + (q_t(s) )^{-\varrho -\varsigma }\expon{\ve\Abs{\ln \left(\frac1{\varphi_t}+\frac1\PF\right)}}\right)\right)\I{s\in \mathbb S_t}\right.\\
&&\left.\quad + \frac{(q_t(s))^{\ve-\varsigma }}{(1-\ve)|B(t_0-1)|}
\left(\frac{(q_t(s))^{\ve-\varrho }}{(1-\ve)|\tilde A(t_0-1)|} (2t_0^{\alpha }/L(\IF)+1) + \frac{(1/\varphi_t+1/q_t(s))^{\varrho}-1}{\varrho}\right)\I{s\in \mathbb S_t^c}
\right],
\EQNY
which is integrable in $[0,1]$.  
Consequently, by the dominated convergence theorem
\BQNY
\lefteqn{
\frac{1}{\tilde A(\varphi_t)} \int_0^1I_{3t}(s)\,ds =  \int_0^1 ({\PF})^{\alpha }\left(1+\frac{\PF}{\varphi_t}\right)^{\kappa-\alpha }
\frac1{\tilde A(\varphi_t)}\left(\frac{L\left(1+\frac{\varphi_t}{\PF}\right)}{L\left(\varphi_t\right)}-1\right)\,ds
}\\
&&=\int_0^1\left(D_\gamma\fracl1s + H_{\gamma, \rho}\fracl1s  A (t) (1+o(1))\right)^{\alpha } \left(1+(\kappa-\alpha )\frac{D_\gamma(1/s)}{\varphi_t}(1+o(1))\right) \\
&&\quad\times \left(\frac{(q_t(s))^{-\varrho } (1+q_t(s)/\varphi_t)^{\varrho } -1}{\varrho } + \frac{(q_t(s))^{-\varrho -\varsigma }  -1}{\varrho +\varsigma } \tilde B(\varphi_t) (1+o(1))\right)\,ds \\
&&=N_{\alpha, 0, \varrho}+A (t)  N_{\alpha, 1, \varrho}  (1+o(1)) + \frac{(\kappa-\alpha )N_{\alpha+1, 0, \varrho}+ L_{\alpha -\varrho +1}}{\varphi_t} (1+o(1)) +\tilde B(\varphi_t) N_{\alpha, 0,\varrho +\varsigma} (1+o(1))
\EQNY
which together with \eqref{decomp.1}, \eqref{I1-third} and \eqref{I2-third} establishes the claim in \eqref{Expan.Third.GW}. 
\QED

\proofkorr{Cor1} We adopt the notation as in the proof of \netheo{T2}. A straightforward application \eqref{MEF_decom}   with $S\sim Beta(1,\kappa)$ (note that $\pk{S>1-s}=s^\kappa, s\in(0,1)$), we have with $\alpha=\kappa$
\BQNY
\frac{\E{(X-U(t))_+^\kappa}}{t^{-1} (a(t))^\kappa} =\int_0^1(q_t(s))^\kappa\,ds = L_\kappa +\int_0^1 I_{1t}(s)\,ds.
\EQNY
(i) For $\gamma>0$, it follows that $U\in3\ERV_{\gamma,\rho,\eta}$ with first-, second- and third-order auxiliary functions $a, A$ and $B$ holds with
$H_{\gamma,\rho}$ and $R_{\gamma,\rho,\eta}$ replaced by $H$ and $R$ given by
\BQNY
&&H(x)= \frac{x^\gamma}\gamma\frac{x^\rho-1}\rho=\frac{\gamma+\rho}\gamma H_{\gamma, \rho}(x)+\frac1\gamma D_\gamma(x) \\
&&R(x)= \frac{x^\gamma}\gamma\frac{x^{\rho+\eta}-1}{\rho+\eta} =\frac1\gamma\Big(\eta(\gamma+\rho+\eta)R_{\gamma, \rho,\eta}(x)+ (\gamma+\rho+\eta)H_{\gamma, \rho}(x)+D_\gamma(x)\Big).
\EQNY
Thus, it follows from \eqref{I1-third} and Remark \ref{rem1} that \eqref{expans} holds.
Further, note that for all $c\in\R$ such that   $\kappa(\gamma+c)<1$
\BQNY
\int_0^1\big(D_\gamma(1/s)\big)^\kappa s^{-\gamma c}\,ds=\frac1{\gamma^{\kappa+1}}B\left(\frac1\gamma-\kappa-c, \kappa+1\right).
\EQNY
Consequently the claim follows by cumbersome calculations. \\
(ii) For $\gamma<0$, we have $x_F-U\in 3\ERV_{\gamma, \rho, \eta}$ with first-, second- and third-order auxiliary functions $a, A$ and $B$ and the limit functions $H_{\gamma,\rho}$ and $R_{\gamma,\rho,\eta}$ replaced by $H$ and $R$ as above. Further, for  $ \kappa, c>0$
\BQNY
\int_0^1\big(D_\gamma(1/s)\big)^{\kappa-1}s^{-\gamma c}\,ds=\frac1{(-\gamma)^\kappa }B\left(c-\frac1\gamma, \kappa\right).
\EQNY
Consequently, the claim follows by similar arguments.
\QED

 \prooftheo{T3}  It follows from Corollary \ref{Cor1} that, as $t\to\IF$
\BQN\label{ExpanD}
\lefteqn{\frac{\big(\E{(X-U(t))_+^{\kappa-1}}\big)^\kappa }{\big(\E{(X-U(t))_+^\kappa }\big)^{\kappa-1}}
}\notag\\
&&=\frac1t\frac{\big(L_{\kappa -1} +A(t)(M_{\kappa-1, 1} + A(t)M_{\kappa-1, 2}+B(t)Q_{\kappa -1})(1+o(1))\big)^\kappa }
{\big(L_\kappa  +A(t)(M_{\kappa,1} + A(t)M_{\kappa ,2}+B(t)Q_\kappa )(1+o(1))\big)^{\kappa-1}}\notag\\
&&=\frac{L_{\kappa -1}^\kappa }{tL_\kappa ^{\kappa-1}}\left(1+\left(\kappa\frac{M_{\kappa-1, 1}}{L_{\kappa -1}}  - (\kappa-1)\frac{M_{\kappa,1}}{L_\kappa }\right)A(t) + \left(\kappa\frac{Q_{\kappa -1}}{L_{\kappa -1}}  - (\kappa-1)\frac{Q_\kappa }{L_\kappa }\right)A(t)B(t)(1+o(1)) \right. \notag\\
&&\quad+\left. \left(\kappa\frac{M_{\kappa-1, 2}}{L_{\kappa -1}}- (\kappa-1)\frac{M_{\kappa ,2}}{L_\kappa } +\frac{\kappa(\kappa-1)}2\left(\frac{M_{\kappa,1}}{L_\kappa }-\frac{M_{\kappa-1, 1}}{L_{\kappa -1}} \right)^2\right)A^2(t) (1+o(1))\right) \notag\\
&&=: \frac \c t\Big(1+\Delta_\kappa  A(t) +\big(\Theta_\kappa  A(t)B(t)  +\Lambda_\kappa  A^2(t)\big)(1+o(1))\Big)
\EQN
with
\BQNY
&&\c=\frac{L_{\kappa -1}^\kappa }{L_\kappa ^{\kappa-1}}, \quad\Delta_\kappa =\kappa\frac{M_{\kappa-1, 1}}{L_{\kappa -1}}  - (\kappa-1)\frac{M_{\kappa,1}}{L_\kappa }, \quad \Theta_\kappa  =\kappa\frac{Q_{\kappa -1}}{L_{\kappa -1}}  - (\kappa-1)\frac{Q_\kappa }{L_\kappa } \\
&&\Lambda_\kappa =\kappa\frac{M_{\kappa-1, 2}}{L_{\kappa -1}}- (\kappa-1)\frac{M_{\kappa ,2}}{L_\kappa } +\frac{\kappa(\kappa-1)}2\left(\frac{M_{\kappa,1}}{L_\kappa }-\frac{M_{\kappa-1, 1}}{L_{\kappa -1}} \right)^2.
\EQNY
Note that (see also \eqref{ERV_12}) $A\in2\RV_{\rho,\eta}$ with auxiliary function $B$ and $\abs{B}\in\RV_{\eta}$.
We have by \eqref{Eq: kq} and \eqref{ExpanD} that $(1-q)t =\c(1+\Delta_\kappa  A(t)(1+o(1)))$. Further, with $\epsilon_q=A(1/(1-q)), \psi_q=B(1/(1-q))$, as $q\uparrow1$
\BQN\label{ExpanAB}
B(t) &=&  \c^\eta \psi_q(1+o(1))\notag\\
A(t)&=&\epsilon_q\Big(c(1+\Delta_\kappa  A(t)(1+o(1)))\Big)^\rho\left(1+\frac{ \c^\eta-1}\eta B\cL{\fracl1{1-q}}(1+o(1))\right) \notag \\
&=& \c^\rho\epsilon_q +\rho  \c^{2\rho}\Delta_\kappa  \epsilon_q^2(1+o(1))  + \c^{\cL{\rho}} \frac{ \c^\eta-1}\eta \epsilon_q\psi_q(1+o(1))
\EQN
Therefore, it follows by \eqref{ExpanD} and \eqref{ExpanAB} that, the solution $t=t(q)$ to \eqref{Eq: kq} has the following third-order expansion
\BQN\label{(1-q)t}
t=\frac \c{1-q}\left(1+ \c^\rho \Delta_\kappa  \epsilon_q+ \c^{2\rho}(\rho \Delta^2_\kappa +\Lambda_\kappa ) \epsilon_q^2(1+o(1))+ \c^{\cL{\rho}}\left(\cL{\c^\eta}\Theta_\kappa +\Delta_\kappa \frac{ \c^\eta-1}{\eta}\right) \epsilon_q\psi_q(1+o(1))\right).
\EQN
Thus, by \eqref{Eq: kq}, Corollary \ref{Cor1} and \eqref{ExpanAB}
\BQN\label{HG2}
\lefteqn{\fracl{\E{(X-U(t))^\kappa _+}}{1-q}^{1/\kappa} = \frac{\E{(X-U(t))^\kappa _+}}{\E{(X-U(t))^{\kappa-1}_+}}}\notag\\
&&= a(t)\frac{ L_\kappa  + A(t)M_{\kappa,1}  +  A(t)\left(M_{\kappa ,2} A(t) + Q_\kappa B(t)\right)(1+o(1))}
{L_{\kappa -1} + A(t) M_{\kappa-1, 1} + A(t)\left(M_{\kappa-1, 2} A(t) + Q_{\kappa -1}B(t)\right) (1+o(1))} \notag\\
&&= a(t)\frac{L_\kappa }{L_{\kappa -1}}\left(1+ \left(\frac{M_{\kappa,1}}{L_\kappa } - \frac{M_{\kappa-1, 1}}{L_{\kappa -1}}\right)A(t) +\left(\frac{Q_\kappa }{L_\kappa } - \frac{Q_{\kappa -1}}{L_{\kappa -1}}\right)A(t)B(t) (1+o(1))\right. \notag\\
&&\quad+\left.  \left(\frac{M_{\kappa ,2}}{L_\kappa }  - \frac{M_{\kappa-1, 2}}{L_{\kappa -1}} +\frac{M_{\kappa-1, 1}^2}{L_{\kappa -1}^2} - \frac{M_{\kappa,1}}{L_\kappa }\frac{M_{\kappa-1, 1}}{L_{\kappa -1}} \right)A^2(t) (1+o(1))\right)\notag\\
&&=a(t)\frac{L_\kappa }{L_{\kappa -1}}\left(1+\epsilon_q \c^\rho\tilde M_{\kappa,1} +\epsilon_q^2 \c^{2\rho}\big[\tilde M_{\kappa ,2}+\rho\tilde M_{\kappa,1}\Delta_\kappa \big](1+o(1)) +\epsilon_q\psi_q \c^{\cL{\rho}}\left[\cL{\c^\eta}\tilde Q_\kappa +\tilde M_{\kappa,1}\frac{ \c^\eta-1}\eta\right](1+o(1))
\right)\notag\\
\EQN
with
\BQNY
&&\tilde M_{\kappa,1}=\frac{M_{\kappa,1}}{L_\kappa } - \frac{M_{\kappa-1, 1}}{L_{\kappa -1}}, \quad \tilde Q_\kappa =\frac{Q_\kappa }{L_\kappa } - \frac{Q_{\kappa -1}}{L_{\kappa -1}} \\
&&\tilde M_{\kappa ,2}=\frac{M_{\kappa ,2}}{L_\kappa }  - \frac{M_{\kappa-1, 2}}{L_{\kappa -1}} +\frac{M_{\kappa-1, 1}^2}{L_{\kappa -1}^2} - \frac{M_{\kappa,1}}{L_\kappa }\frac{M_{\kappa-1, 1}}{L_{\kappa -1}}.
\EQNY
 We will present next the proof for $\gamma>0$ and $\rho, \eta<0$. The other cases follow by similar arguments and thus are omitted here.
 \\
Since $U\in3\RV_{\gamma,\rho,\eta}$ with auxiliary functions $A$ and $B$, we have $U\in3\ERV_{\gamma,\rho,\eta}$ with first-, second- and third-order auxiliary functions $a(t)=\gamma U(t), A(t)$ and $B(t)$. By \nelem{L1} and Remark \ref{rem1}, we have using \eqref{(1-q)t} with $t=t_q$, as $q\uparrow1$
\BQN\label{HG1}
\notag
\lefteqn{\frac{U(t)}{U(1/(1-q))}}\\
 &&= ((1-q)t)^\gamma + \epsilon_q ((1-q)t)^\gamma \frac{((1-q)t)^\rho-1}\rho + \epsilon_q\psi_q((1-q)t)^\gamma \frac{((1-q)t)^{\rho+\eta}-1}{\rho+\eta} (1+o(1))\notag\\
&&=  \c^\gamma \left(1+\gamma  \c^\rho \Delta_\kappa  \epsilon_q+\gamma  \c^{2\rho}\left(\Lambda_\kappa  +\left[\rho+\frac{\gamma-1}2\right] \Delta^2_\kappa \right) \epsilon_q^2(1+o(1))+\gamma  \c^{\cL{\rho}}\left(\cL{\c^\eta}\Theta_\kappa +\Delta_\kappa \frac{ \c^\eta-1}{\eta}\right) \epsilon_q\psi_q(1+o(1))\right)\notag \\
&&\quad +\epsilon_q  \c^\gamma\frac{ \c^\rho-1}\rho\left(1+\left[\gamma  \c^\rho \Delta_\kappa  +\frac{\rho  \c^{2\rho}}{ \c^\rho-1}\Delta_\kappa \right] \epsilon_q (1+o(1))\right) + \epsilon_q\psi_q  \c^\gamma \frac{ \c^{\rho+\eta}-1}{\rho+\eta}(1+o(1))\notag\\
&&= \c^\gamma\left(1+ \epsilon_q\left[\gamma  \c^\rho \Delta_\kappa +\frac{ \c^\rho-1}\rho\right] +  \epsilon_q\psi_q\left[\gamma  \c^{\cL{\rho}}\left(\cL{\c^\eta}\Theta_\kappa +\Delta_\kappa \frac{ \c^\eta-1}{\eta}\right)+\frac{ \c^{\rho+\eta}-1}{\rho+\eta} \right](1+o(1))\right. \notag\\
&&\quad \left. + \epsilon^2_q\left[\gamma  \c^{2\rho}\left(\Lambda_\kappa  +\left[\rho+\frac{\gamma-1}2\right] \Delta^2_\kappa \right)+\frac{ \c^\rho-1}\rho\left(\gamma  \c^\rho \Delta_\kappa  +\frac{\rho  \c^{2\rho}}{ \c^\rho-1}\Delta_\kappa \right) \right](1+o(1))\right),
\EQN
which together with \eqref{HG2} yields that (recall $a(t)=\gamma U(t)$)
\BQN\label{HF}
\notag H_q[X] &=& U(t) +\fracl{\E{(X-U(t))^\kappa _+}}{1-q}^{1/\kappa} \\
\notag &=&  \c^\gamma F^\leftarrow(q) \left(1+\frac{\gamma L_\kappa }{L_{\kappa -1}}\right)
\left(1+\epsilon_q\frac{\gamma L_\kappa }{L_{\kappa -1}} \left(1+\frac{\gamma L_\kappa }{L_{\kappa -1}}\right)^{-1}\right.\\
\notag && \left. \times \left( \c^\rho\tilde M_{\kappa,1} +\epsilon_q \c^{2\rho}[\tilde M_{\kappa ,2}+\rho\tilde M_{\kappa,1}\Delta_\kappa ](1+o(1)) +\psi_q \c^{\cL{\rho}}\left[\cL{\c^\eta}\tilde Q_\kappa +\tilde M_{\kappa,1}\frac{ \c^\eta-1}\eta\right](1+o(1))\right)
\right)\\
\notag &&  \times \left(1+ \epsilon_q\left[\gamma  \c^\rho \Delta_\kappa +\frac{ \c^\rho-1}\rho\right] +  \epsilon_q\psi_q\left[\gamma  \c^{\cL{\rho}}\left(\cL{\c^\eta}\Theta_\kappa +\Delta_\kappa \frac{ \c^\eta-1}{\eta}\right)+\frac{ \c^{\rho+\eta}-1}{\rho+\eta} \right](1+o(1))\right. \notag\\
\notag && \left. + \epsilon^2_q\left[\gamma  \c^{2\rho}\left(\Lambda_\kappa  +\left[\rho+\frac{\gamma-1}2\right] \Delta^2_\kappa \right)+\frac{ \c^\rho-1}\rho\left(\gamma  \c^\rho \Delta_\kappa  +\frac{\rho  \c^{2\rho}}{ \c^\rho-1}\Delta_\kappa \right) \right](1+o(1))\right) \\
&=:& c_0 F^\leftarrow(q)\Big(1+c_1 \epsilon_q +c_2\epsilon_q^2(1+o(1)) +c_3\epsilon_q\psi_q (1+o(1))\Big),
\EQN
which together with ${L_\kappa }/{L_{\kappa -1}}= \kappa/{(1-\kappa\gamma)}$ implies that
\BQNY
c_0 &=& \frac{\c^\gamma}{1-\kappa\gamma}, \quad c_1=\gamma \c^\rho (\kappa\tilde M_{\kappa,1} + \Delta_\kappa )+\frac{ \c^\rho-1}\rho \\
c_2&=&  \gamma  \c^{2\rho}[\kappa\tilde M_{\kappa ,2}+\rho \kappa\tilde M_{\kappa,1}\Delta_\kappa ]
+ \gamma  \c^\rho \kappa\tilde M_{\kappa,1} \left[\gamma  \c^\rho \Delta_\kappa  +\frac{ \c^\rho-1}\rho\right] \\
&&+\gamma  \c^{2\rho}\left(\Lambda_\kappa  +\left[\rho+\frac{\gamma-1}2\right] \Delta^2_\kappa \right)+\frac{ \c^\rho-1}\rho\left(\gamma  \c^\rho \Delta_\kappa  +\frac{\rho  \c^{2\rho}}{ \c^\rho-1}\Delta_\kappa \right) \\
c_3&=&  \kappa\gamma
 \c^{\cL{\rho}}\left[\cL{\c^\eta}\tilde Q_\kappa +\tilde M_{\kappa,1}\frac{ \c^\eta-1}\eta\right]
+\gamma
 \c^{\cL{\rho}}\left(\cL{\c^\eta}\Theta_\kappa +\Delta_\kappa \frac{ \c^\eta-1}{\eta}\right)
+\frac{ \c^{\rho+\eta}-1}{\rho+\eta}
\EQNY
Next, we will calculate  the four coefficients $c_0,\ldots, c_3$ in turn. \\
Recalling that $\c=L_{\kappa-1}^\kappa/ L_\kappa^{\kappa-1}$, we have 
\BQNY \c=\kappa\fracl{1-\kappa\gamma}{\kappa|\gamma|}^\kappa \xi_{\kappa ,0}, \quad \xi_{\kappa ,\rho}=\frac{(\kappa-1)|\gamma|}{1-\rho-\kappa\gamma}\xi_{\kappa -1,\rho}.
\EQNY
Further, by \eqref{ExpanD} and \eqref{HG2} (recall the symmetry of $(\tilde M_{\kappa,1}, \Delta_\kappa )$
and $(\tilde Q_\kappa , \Theta_\kappa )$), we have that \eqref{def: xi} holds and further
\BQN
\label{cal_M}
\left\{\begin{array}{l}
\kappa\tilde M_{\kappa,1}+\Delta_\kappa  = \frac{M_{\kappa,1}}{L_\kappa } = \frac1{\gamma\rho}\left(\frac{\xi_{\kappa ,\rho}}{\xi_{\kappa ,0}}-1\right)\\
\kappa\tilde Q_\kappa +\Theta_\kappa =\frac{Q_\kappa }{L_\kappa }=\frac1{\gamma(\rho+\eta)}\left(\frac{\xi_{\kappa ,\rho+\eta}}{\xi_{\kappa ,0}}-1\right) \\
 \tilde Q_\kappa  = \frac1{\gamma(\rho+\eta)}\left(\frac{\xi_{\kappa ,\rho+\eta}}{\xi_{\kappa ,0}} - \frac{\xi_{\kappa -1,\rho+\eta}}{\xi_{\kappa -1,0}}\right)\\
\frac{M_{\kappa ,2}}{L_\kappa }= \frac1{2\gamma^2\rho^2}\left((1-\gamma-2\rho)\frac{\xi_{\kappa ,2\rho}}{\xi_{\kappa ,0}}-2(1-\gamma-\rho)\frac{\xi_{\kappa ,\rho}}{\xi_{\kappa ,0}}+1-\gamma\right).
\end{array}
\right.
\EQN
Therefore
\BQNY
&&\Theta_\kappa =\frac1{\gamma(\rho+\eta)}\left(\kappa\frac{\xi_{\kappa -1,\rho+\eta}}{\xi_{\kappa -1,0}}
- (\kappa-1)\frac{\xi_{\kappa ,\rho+\eta}}{\xi_{\kappa ,0}}-1\right).
\EQNY
Hence,  by \eqref{cal_M}
\BQNY
c_1=   \c^\rho\left(\gamma\frac{M_{\kappa,1}}{L_\kappa }+\frac1\rho\right)-\frac1\rho = \frac1\rho\left( \c^\rho\frac{\xi_{\kappa ,\rho}}{\xi_{\kappa ,0}}-1\right).
\EQNY
Moreover, we rewrite $c_2$ as follows.
\BQNY
c_2=  \gamma  \c^{2\rho} \left[\kappa\tilde M_{\kappa ,2} +\Lambda_\kappa  +(\gamma+\rho)\kappa\tilde M_{\kappa,1}\Delta_\kappa  +\left(\rho+\frac{\gamma-1}2\right)\Delta_\kappa ^2+\frac1\gamma \Delta_\kappa \right]  + \gamma  \c^\rho\frac{ \c^\rho-1}\rho(\kappa\tilde M_{\kappa,1}+\Delta_\kappa ).
\EQNY
Note further by \eqref{ExpanD} and \eqref{HG2} that
\BQNY
\kappa\tilde M_{\kappa ,2} +\Lambda_\kappa =\frac{M_{\kappa ,2}}{L_\kappa } +\kappa\tilde M_{\kappa,1}\left(\frac{\kappa-1}2\tilde M_{\kappa,1}-\frac{M_{\kappa-1, 1}}{L_{\kappa -1}}\right).
\EQNY
The claim for $c_2$ follows by \eqref{cal_M}. \\
Finally, it follows again  by \eqref{cal_M} that
\BQNY c_3
&=& \c^{\rho+\eta}\left[\gamma \kappa\tilde
Q_\kappa +\gamma\Theta_\kappa +\frac1{\rho+\eta}\right] - \frac{1}{\rho+\eta} +
\gamma  \c^{\cL{\rho}}\frac{ \c^\eta-1}{\eta}\frac{M_{\kappa,1}}{L_\kappa } \\
&=&\frac1{\rho+\eta}\left( \c^{\rho+\eta}\frac{\xi_{\kappa ,\rho+\eta}}{\xi_{\kappa ,0}}-1\right)
+\frac{ \c^{\cL{\rho}}( \c^\eta-1)}{\rho\eta}\left(\frac{\xi_{\kappa ,\rho}}{\xi_{\kappa ,0}}-1\right).
 \EQNY 
  Now for $\gamma<0$, note that $a(t)=-\gamma(x_F-U(t))$  and $x_F-U(t)\in 3\RV_{\gamma, \rho, \eta}$.  It follows that \eqref{HG1} holds for $(x_F-U(t))/(x_F-U(1/(1-q)))$. Further \eqref{HF} holds by replacing $H_q[X]$ and $F^\leftarrow(q)$ by $x_F-H_q[X]$ and $x_F-F^\leftarrow(q)$, respectively. The remaing arguments are the same as for $\gamma>0$. This completes the proof of \netheo{T3}.
 \QED

\prooftheo{T4} Since $\alpha:=1/\gamma>1$, we have $\E X<\IF$ and thus the expectile $e_q$ satisfies \eqref{def: expectile}. It follows by Theorem 2.3.9 in \cite{deHaanF2006} that $\overline F\in 2\RV_{-\alpha, \alpha\rho}$ with auxiliary function $A^*(x)= \alpha^2 A(1/\overline F(x))$. By Corollary 4.4 in \cite{MaoH2012}
\BQNY
\E{(X-x)_+} = \frac{x \overline F(x)}{\alpha-1}\left(1+\frac{A^*(x)}{\alpha-1-\alpha\rho}(1+o(1))\right),\quad  x\to\IF,
\EQNY
which together with the first-order approximation $e_q=(\alpha-1)^{-1/\alpha} F^\leftarrow(q) (1+o(1)) \to\IF, q\uparrow1$ (see Proposition 2.3 in \cite{BelliniB2014})
yields that
\BQN\label{first_qF}
e_q&=& \frac{1}{1-\E{X}/e_q}\frac{2q-1}{1-q}\E{(X-e_q)_+}\notag \\
&=& \frac{e_q}{1-q}\frac{\overline F(e_q) }{\alpha-1}(1-2(1-q))\left(1+\frac{A^*(e_q)}{\alpha-1-\alpha\rho}(1+o(1)) + \frac{ (\alpha-1)^{1/\alpha}\E{X}}{F^\leftarrow(q)}(1+o(1))\right)\notag\\
&=&\frac{e_q}{1-q} \frac{\overline F(e_q)}{\alpha-1}
\left(1+  \frac{\alpha d_0}{F^\leftarrow(q)}(1+o(1))+\alpha d_1(1-q)(1+o(1)) +\alpha D \epsilon_q(1+o(1))\right).
\EQN
The last step follows since $|A|\in \RV_\rho$ and $(1-q)/\overline F(e_q)=1/(\alpha-1)(1+o(1))$.
\\
Further by $U\in2\RV_{\gamma, \rho}$ with auxiliary function $A$
\BQNY
e_q &=& U\left(\frac1{1-q} \frac{1}{\alpha-1} \left(1+  \frac{\alpha d_0}{F^\leftarrow(q)}(1+o(1))+\alpha d_1(1-q)(1+o(1)) +\alpha D \epsilon_q(1+o(1))\right)\right)\\
      &=& \overline F^\leftarrow(q)\left[\frac{1}{\alpha-1} \left(1+  \frac{\alpha d_0}{F^\leftarrow(q)}(1+o(1))+\alpha d_1(1-q)(1+o(1)) +\alpha D \epsilon_q(1+o(1))\right)\right]^{1/\alpha}\\
      &&\times \left[1+\frac{(\alpha-1)^{-\rho}-1}{\rho}\epsilon_q(1+o(1))\right]\\
      &=&  (\alpha-1)^{-1/\alpha} F^\leftarrow(q)  \left(1+  \frac{d_0}{F^\leftarrow(q)}(1+o(1))+ d_1(1-q)(1+o(1)) + d_2 \epsilon_q(1+o(1))\right)
\EQNY
establishing the proof of the first claim.\\
First by \nelem{L1} \BQNY \E{(X-x)_+} = \frac{x \overline
F(x)}{\alpha-1}\left(1+\frac{A^*(x)}{\alpha-1-\alpha\rho}\left(
1+\frac{\alpha-1-\alpha\rho}{\alpha-1-\alpha(\rho+\eta)}B^*(x)(1+o(1))\right)\right)
\EQNY with $A^*(x)=\alpha^2A(1/\overline F(x)), B^*(x)=
B(1/\overline F(x))$. Therefore \BQNY
\frac1{\overline F(e_q)}&=& \frac{1-2(1-q)}{1-q}\frac{1+\E{X}/e_q+(\E{X}/e_q)^2(1+o(1))}{\alpha-1}\notag \\
&&\times \left[1+
\frac{A^*(e_q)}{\alpha-1-\alpha\rho}\left( 1+\frac{\alpha-1-\alpha\rho}{\alpha-1-\alpha(\rho+\eta)}B^*(e_q)(1+o(1))\right)\right]\notag \\
&=&\frac{1}{1-q}\frac1{\alpha-1}\left[1-2(1-q) + \frac{\E{X}}{e_q} +\frac{A^*(e_q)}{\alpha-1-\alpha\rho} \right. \notag\\
&&-2(1-q)\frac{\E{X}}{e_q}(1+o(1))-2(1-q)\frac{A^*(e_q)}{\alpha-1-\alpha\rho}(1+o(1)) + \frac{A^*(e_q)}{\alpha-1-\alpha\rho}\frac{\E{X}}{e_q}(1+o(1))\notag\\
&&\left. + \fracl{\E{X}}{e_q}^2(1+o(1))+\frac{A^*(e_q)B^*(e_q)}{\alpha-1-\alpha(\rho+\eta)}(1+o(1))\right]\notag\\
&=:&\frac{x_q}{1-q}.
\EQNY
Noting that $A\in2\RV_{\rho, \eta}$ with auxiliary function $B$, and $\abs{B}\in\RV_\eta$, it follows by \eqref{first_qF} that
\BQNY
\frac{A^*(e_q)}{\alpha-1-\alpha\rho}&=&\frac{\alpha^2}{\alpha-1-\alpha\rho}\epsilon_q x_q^\rho \left(1+\frac{x_q^\eta-1}{\eta}\psi_q(1+o(1))\right) = \alpha D\epsilon_q\left[1-2\rho(1-q) (1+o(1)) \right.\\
&&\left.+\frac{\alpha\rho d_0}{F^\leftarrow(q)}(1+o(1)) + \alpha\rho D\epsilon_q (1+o(1))+\frac{(\alpha-1)^{-\eta}-1}\eta\psi_q(1+o(1))\right]\\
B^*(e_q)&=& (\alpha-1)^{-\eta}\psi_q(1+o(1)),
\EQNY
which together with \netheo{T3} implies that
\BQN\label{def: xq}
x_q&=& \frac1{\alpha-1}\left[1-2(1-q) +  \frac{\alpha d_0}{F^\leftarrow(q)} \left(1- \frac{d_0}{F^\leftarrow(q)}(1+o(1)) - d_1(1-q)(1+o(1)) -d_2 \epsilon_q(1+o(1))\right) \right.\notag \\
&&+
\alpha  D\epsilon_q\left( 1-2\rho(1-q)(1+o(1))+ \frac{\alpha\rho d_0}{F^\leftarrow(q)}(1+o(1)) + \alpha\rho D\epsilon_q (1+o(1))+\frac{(\alpha-1)^{-\eta}-1}\eta\psi_q(1+o(1))\right)\notag\\
&& -\alpha\left(2 d_0\frac{1-q}{F^\leftarrow(q)}+2 D(1-q)\epsilon_q-\alpha D d_0 \frac{\epsilon_q}{F^\leftarrow(q)} \left.-\frac{\alpha d_0^2}{(F^\leftarrow(q))^2} -\frac{\alpha(\alpha-1)^{-(\rho+\eta)}}{\alpha-1-\alpha(\rho+\eta)}\epsilon_q\psi_q\right)(1+o(1)) \right].
\EQN
Next, by $U\in3\RV_{\gamma, \rho, \eta}$ with auxiliary functions $A$ and $B$
\BQN\label{eq-third}
\frac{e_q}{F^\leftarrow(q)}= \frac{U(x_q/(1-q))}{U(1/(1-q))} = x_q^{1/\alpha}\left(1+\frac{x_q^\rho-1}\rho \epsilon_q +\frac{x_q^{\rho+\eta}-1}{\rho+\eta} \epsilon_q\psi_q(1+o(1))\right),
\EQN
with (recall $x_q$ defined in \eqref{def: xq})
\BQNY
x_q^{1/\alpha}&=& (\alpha-1)^{-1/\alpha}\left[1+d_1(1-q) +   \frac{d_0}{F^\leftarrow(q)} \left(1- \frac{d_0}{F^\leftarrow(q)}(1+o(1)) - d_1(1-q)(1+o(1)) -d_2 \epsilon_q(1+o(1))\right) \right. \\
&&+
D\epsilon_q\left( 1-2\rho(1-q)(1+o(1))+ \frac{\alpha\rho d_0}{F^\leftarrow(q)}(1+o(1)) + \alpha\rho D\epsilon_q (1+o(1))+\frac{(\alpha-1)^{-\eta}-1}\eta\psi_q(1+o(1))\right)\\
&& -\left(2 d_0\frac{1-q}{F^\leftarrow(q)}+2 D(1-q)\epsilon_q-\alpha D d_0 \frac{\epsilon_q}{F^\leftarrow(q)} -\frac{\alpha d_0^2}{(F^\leftarrow(q))^2} -\frac{\alpha(\alpha-1)^{-(\rho+\eta)}}{\alpha-1-\alpha(\rho+\eta)}\epsilon_q\psi_q\right)(1+o(1)) \\
&&+\left.\frac{1-\alpha}{2}\left(\frac{d_0^2}{(F^\leftarrow(q))^2}+d_1^2(1-q)^2 +D^2\epsilon_q^2\right)(1+o(1))
\right]\\
\frac{x_q^\rho-1}\rho&=& \frac{1}{\rho}\left( (\alpha-1)^{-\rho}\left[1+\alpha\rho d_1(1-q)(1+o(1)) +\frac{\alpha\rho d_0}{F^\leftarrow(q)}(1+o(1))+ \alpha\rho D\epsilon_q(1+o(1))\right]-1\right)\\
\frac{x_q^{\rho+\eta}-1}{\rho+\eta} &=&
\frac{(\alpha-1)^{-(\rho+\eta)}-1}{\rho+\eta}(1+o(1)). \EQNY
Consequently, the desired result follows by \eqref{eq-third} and
elementary calculations. \QED

\prooftheo{T7} By Corollary 4.4 in \cite{MaoH2012} \BQNY \E{(X-x)_+}
= \frac{x_F-x}{\alpha+1} \overline
F(x)\left(1-\frac{\alpha^2}{\alpha+1-\alpha\rho}A\fracl1{\overline
F(x)}(1+o(1))\right), \quad x\uparrow x_F. \EQNY Further by
\eqref{def: expectile} \BQNY e_q - \E{X} = \frac{x_F-e_q}{1-q}\frac{
\overline F(e_q)}{\alpha+1}\left( 1-2(1-q)(1+o(1)) -
\frac{\alpha^2}{\alpha+1-\alpha\rho}A\fracl1{\overline
F(e_q)}(1+o(1))\right). \EQNY It follows from Proposition 2.5 in
\cite{BelliniB2014} that $x_F-e_q = \widetilde C
(1-q)^{1/(\alpha+1)}\to0$ for some positive constant $\widetilde C$.
We have thus $t_q:=(x_F-e_q)/(1-q)\to\IF$. By \eqref{def: expectile}
and Taylor's expansion $1/(1-x)= 1+x(1+o(1)), \, x\to0$ \BQNY
\frac1{\overline F(e_q)} &=& \frac{x_F-e_q}{1-q}\frac{1}{(\alpha+1)(x_F-\E{X})}\\
&&\times \left( 1- \frac{\alpha^2}{\alpha+1-\alpha\rho}A\fracl1{\overline F(e_q)}(1+o(1)) + \frac{x_F-e_q}{x_F-\E{X}}(1+o(1))\right)\\
&=:& \frac{t_q}{x_0}\left( 1-
\frac{\alpha^2}{\alpha+1-\alpha\rho}A\fracl1{\overline
F(e_q)}(1+o(1)) + \frac{x_F-e_q}{x_F-\E{X}}(1+o(1))\right). \EQNY
Noting further that $x_F-U\in 2\RV_{\gamma, \rho}$ with auxiliary
function $A$ and $|A|\in\RV_\rho$ \BQN\label{endp-e}
x_F - e_q &=& (x_F-U(t_q))x_0^{1/\alpha}\left(1+\frac{x_0^{-\rho}-1}\rho A(t_q)(1+o(1))\right)\notag\\
&&\times \left( 1 + \frac{\alpha}{\alpha+1-\alpha\rho}A\fracl1{\overline F(e_q)}(1+o(1))+\frac{\gamma(x_F-e_q)}{x_F-\E{X}}(1+o(1))\right) \notag\\
&=& C\fracl{x_0}{t_q}^{1/\alpha} \left(1+\frac{x_0^{-\rho}}\rho A(t_q)(1+o(1)) + \frac{\alpha}{\alpha+1-\alpha\rho}A\fracl1{\overline F(e_q)}(1+o(1)) +\frac{\gamma(x_F-e_q)}{x_F-\E{X}}(1+o(1))\right)\notag \\
&=&
C\fracl{x_0}{t_q}^{1/\alpha}\left(1+\frac{(\alpha+1)x_0^{-\rho}}{\rho(\alpha+1-\alpha\rho)}A(t_q)(1+o(1))
-\frac{x_F-e_q}{\alpha(x_F-\E{X})}(1+o(1))\right). \EQN Clearly,
$x_F-e_q= C(x_0/t_q)^{1/\alpha}(1+o(1))$ we have (recall
$t_q:=(x_F-e_q)/(1-q)$) \BQNY x_F-e_q =
(C^{\alpha}x_0(1-q))^{1/(\alpha+1)} (1+o(1)) \EQNY and thus \BQNY
t_q= \fracl C{1-q}^{\alpha/(\alpha+1)}x_0^{1/(\alpha+1)} (1+o(1)).
\EQNY Consequently, by \eqref{endp-e} \BQNY
x_F - e_q &=&(C^\alpha x_0(1-q))^{1/(\alpha+1)}\left(1-\frac{(C^\alpha x_0(1-q))^{1/(\alpha+1)}}{\alpha(x_F-\E{X})}(1+o(1)) \right.\\
&&\left.+\frac{(\alpha+1)(C/x_0)^{\alpha\rho/(\alpha+1)}}{\rho(\alpha+1-\alpha\rho)}A\left((1-q)^{-\frac\alpha{\alpha+1}}\right)(1+o(1))\right).
\EQNY We obtain the desired result. \QED

 {\bf Acknowledgements}. The first author acknowledges partial support by the Swiss
National Science Foundation grant 200021-140633/1 and RARE -318984 (an FP7 Marie Curie IRSES Fellowship).
The second author was  supported by the National Natural Science Foundation of China grant 11171275.

\section{Appendix}\label{sec6}
In this appendix we first establish an \emph{extensional Drees' type inequality} in
\nelem{L1} for the third-order extended regularly varying functions.
Then we present a proposition concerning the third-order regular
variation properties under generalized inverse transformations. Recall that $D_\gamma, H_{\gamma, \rho}$ and $R_{\gamma, \rho, \eta}$ are defined by \eqref{ThirdLimit}. 
\BL\label{L1} 
\COM{Let $f$ be
a measurable function. If there exist  some parameters $\alpha\in\R$
and $\rho, \eta\le 0$, and some functions $a_0(\cdot)>0$ and
$a_i(\cdot), i=1, 2$ with constant sign at infinity and $\limit t
a_i(t)=0, i=1,2$, such that for all $x>0$ \BQN\label{ThirdLimit}
\limit t\frac{f(tx)-f(t) - a_0(t) D_\gamma(x) - a_1(t) H_{\gamma,
\rho}(x)}{a_2(t)} = R_{\gamma, \rho, \eta}(x), 
\EQN
}
If $f\in 3\ERV_{\gamma, \rho, \eta}$ with auxiliary functions $a, A$ and $B$, then for any
$\epsilon>0$, there exists $t_0= t_0(\ve)>0, C>0$ such that for all $\min(t, tx)\ge
t_0$ 
\BQN\label{ThirdLimit2} 
\lefteqn{\Abs{\frac{f(tx)-f(t) - a_0(t)
D_\gamma(x) - a_1(t) H_{\gamma, \rho}(x)}{a_2(t)}- R_{\gamma, \rho,
\eta}(x) }}\notag\\
&& \le \epsilon (1+x^\gamma + 2x^{\gamma+\rho} + 4
x^{\gamma+\rho+\eta} e^{\epsilon \abs{\ln x}} \cL{+\I{\gamma=\rho=0}e^{ C\abs{\ln x}}}) 
\EQN
with $a_0(t)=a(t), a_1(t)= a_0(t)A(t)$ and $a_2(t)= a_0(t)A(t)B(t)$. 
\EL
{\remark\label{rem1} a) We see that 
\eqref{ThirdLimit2} also holds for $f\in 3\RV_{\gamma, \rho, \eta}$ with $H_{\gamma, \rho}$ and $R_{\gamma,
\rho, \eta}$  replaced respectively by $H$ and $R$ given by \BQNY
H(x) = c_1H_{\gamma, \rho}(x) +c_2 D_\gamma(x),\quad
R(x)=d_1R_{\gamma, \rho, \eta}(x)+ d_2 H_{\gamma, \rho}(x) +d_3
D_\gamma(x) \EQNY
with $c_i, d_j\in\R, i=1,2, j=1,2,3.$ \\
b) 
The inequality \eqref{ThirdLimit2} is the third-order form of Lemma 5.2 in \cite{DraismadPP1999}, which is the so-called the \emph{extensional Drees\rq{} inequality}, which is different from those by Theorem 2.1 in \cite{FragaAlvesdL2006} and Lemma 2.1 in \cite{WangC2006}
}

\prooflem{L1} For simplicity, we denote
\BQNY
&&I_0(t,x)=\frac{a_0(tx)-a_0(t)x^\gamma - a_1(t) x^\gamma D_\rho(x)}{a_2(t)},\quad I_1(t,x)= \frac{a_1(tx)-a_1(t)x^{\gamma+\rho}}{a_2(t)}\\
&&I_2(t,x)=\frac{a_2(tx)}{a_2(t)},\quad I(t,x)=\frac{f(tx)-f(t) -
a_0(t) D_\gamma(x) - a_1(t) H_{\gamma, \rho}(x)}{a_2(t)}.
\EQNY
It follows from Theorem 2.1 in \cite{FragaAlvesdL2006} that
\BQN\label{ERV_12}
\limit t I_0(t,x)= x^\gamma H_{\rho, \eta}(x), \quad
\limit t I_1(t,x) =x^{\gamma+\rho} D_\eta(x), \quad
\limit t I_2(t,x) = x^{\gamma+\rho+\eta}.
\EQN
Next, we will consider the following four cases: \underline{Case a: $\gamma\neq 0$ and $\gamma+\rho\neq 0$}; \underline{Case b: $\gamma\neq 0$ and $\gamma+\rho=0$};\\
 \underline{Case c: $\gamma=0$ and $\gamma+\rho\neq0$} and \underline{Case d:  $\gamma=\rho=0$}, respectively.
\\
\underline{Case a: $\gamma\neq 0$ and $\gamma+\rho\neq 0$}. Let $g(t)= f(t)-a_0(t)/\gamma +a_1(t)/(\gamma(\gamma+\rho))$. It follows from  \eqref{ThirdLimit} and \eqref{ERV_12} that, for all $x>0$
\BQNY
\frac{g(tx) -g(t)}{a_2(t)/(\gamma(\gamma+\rho))} &=&\gamma(\gamma+\rho)I(t,x)-(\gamma+\rho) I_0(t,x)+I_1(t,x)\\
&\to&  \gamma(\gamma+\rho) R_{\gamma,\rho,\eta}(x)-(\gamma+\rho) x^\gamma H_{\rho, \eta}(x) + x^{\gamma+\rho} D_\eta(x) \\
&=& D_{\gamma+\rho+\eta}(x), \quad t\to\IF.
\EQNY
Hence, it follows from Lemma 5.2 in \cite{DraismadPP1999} and similar arguments of Lemma 2.1 in \cite{deHaanP1997} that, for any $\epsilon >0$, there exists some $t_0=t_0(\ve) >0$ such that for all $\min(t, tx) \ge t_0$
\BQNY
\Abs{\frac{g(tx) -g(t)}{a_2(t)/(\gamma(\gamma+\rho))}- D_{\gamma+\rho+\eta}(x)}&\le & \epsilon (1+x^{\gamma+\rho+\eta}e^{\epsilon\abs{\ln x}}) \\
\Abs{x^{-(\gamma+\rho)}I_1(t,x) - D_\eta(x)} &=&\Abs{\frac{(tx)^{-(\gamma+\rho)}a_1(tx)- t^{-(\gamma+\rho)}a_1(t)}{t^{-(\gamma+\rho)}a_2(t)} - D_\eta(x)} \\
&\le & \epsilon (1+x^\eta e^{\epsilon\abs{\ln x}})
\EQNY
and
\BQNY
\Abs{x^{-\gamma}I_0(t,x) -  H_{\rho, \eta}(x)} &=& \Abs{\frac{(tx)^{-\gamma}a_0(tx)- t^{-\gamma} a_0(t) - t^{-\gamma}a_1(t) D_\rho(x)}{t^{-\gamma}a_2(t)} - H_{\rho, \eta}(x)} \\
&\le & \epsilon (1+x^\rho+ 2x^{\rho+\eta} e^{\epsilon\abs{\ln x}}).
\EQNY
Consequently
\BQNY
\lefteqn{\Abs{I(t,x) - R_{\gamma, \rho, \eta}}}\\
&&\le\epsilon\left( \frac{1+ x^{\gamma+\rho+\eta}e^{\epsilon \abs{\ln x}}}{\gamma(\gamma+\rho)} +
\frac{x^\gamma(1+x^\rho+2x^{\rho+\eta}e^{\epsilon \abs{\ln x}})}\gamma + \frac{x^{\gamma+\rho}(1+x^\eta e^{\epsilon \abs{\ln x}})}{\gamma(\gamma+\rho)} \right)
\EQNY
establishing our proof for Case a. \\
\underline{Case b: $\gamma\neq0$ and $\gamma+\rho=0$}. 
Letting 
$g(t)= f(t)-a_0(t)/\gamma$, we have 
\BQNY
\lefteqn{\frac{g(tx) -g(t)  -(-a_1(t)/\gamma)\ln x}{a_2(t)/\rho}}\\
&=& \rho I(t,x) +I_0(t,x)\\
&\to& \rho R_{\gamma,\rho,\eta}(x) + x^\gamma H_{\rho, \eta}(x)  \\
&=&  H_{0,\eta}(x), \quad t\to\IF.
\EQNY
Consequently, the claim follows by similar arguments for Case a. \\
\underline{Case c: $\gamma=0$ and $\gamma+\rho\neq0$}.
Let $g(t)= f(t) - ({a_1(t) -a_2(t)/(\rho+\eta)})/{\rho^2}$, then
\BQNY
\lefteqn{\frac{g(tx) -g(t)  -(a_0(t)-a_1(t)/\rho)\ln x}{a_2(t)/(-\rho)} }\\
&=& (-\rho)I(t,x)+\frac1\rho I_1(t,x) - \frac1{\rho(\rho+\eta)} (I_2(t,x)-1)\\
& \to& (-\rho)\left(R_{0,\rho,\eta}(x) - \frac{x^\rho}{\rho^2}D_\eta(x) +\frac1{\rho^2}D_{\rho+\eta}(x)\right) \\
&=&  H_{0,\rho+\eta}(x), \quad t\to\IF.
\EQNY
The remaining proof is similar to those for Case a and thus is omitted here.

\underline{Case d: $\gamma=\rho=0$}. We first consider that $\eta<0$. Since $(a_1(tx)-a_1(t))/a_2(t)\to D_\eta(x)$, 
we have by Theorem 1.10 in \cite{Gelukd1987} that,
there exists some constant $c\neq0$ such that
\BQN \label{Lim1}
\limit t a_1(t)=c, \quad \limit t\frac{a_1(t)-c}{a_2(t)} =\frac1\eta.
 \EQN
Letting $g(t)=f(t)-c(\ln t)^2/2$, we have
\BQNY
\frac{g(tx) -g(t)-(a_0(t)-c\ln t) \ln x}{a_1(t)-c} 
&=&\frac{a_2(t)}{a_1(t)-c}\left(I(t,x) - R_{0,0,\eta}(x) + \left(\frac{a_1(t)-c}{a_2(t)}-\frac1\eta\right)\frac{\ln^2x}{2} + \frac1\eta H_{0,\eta}(x)\right)\\
&\to& H_{0,\eta}(x).
\EQNY
We see from \eqref{ThirdLimit} that $g\in 2\ERV_{0,\eta}$ with auxiliary function $a_0(t)-c\ln t$ and $a_1(t)-c$. Noting further that there exist two constants $C, D>0$ such that for all $x\in(0,\IF)$ (cf. Lemma 2.2 in \cite{WangC2006})
\BQN\label{Bound.H}
\cL{\abs{H_{0,\tau}(x)} \le D\expon{C\abs{\ln x}},\quad \tau\le0. }
\EQN
 The claim
follows by \eqref{Lim1} and similar arguments as for Case a.
\\
Next, we deal with the case $\eta=0$. Letting $g(t)= f(t) - \int_1^t{a_0(u)}/u\,du+a_1(t)$, we have
 by \eqref{ThirdLimit} and  \eqref{ERV_12}
\BQNY
\frac{g(tx)-g(t)}{a_2(t)} &=&  I(t,x) -\int_1^x \frac{I_0(t,u)}u\,du +I_1(t,x)\\
&\to& \frac{(\ln x)^3}6 - \int_1^x\frac{(\ln u)^2}{2u}\,du+\ln x=\ln x,
\EQNY
where we used in the second step the dominated convergence  theorem (see e.g. Lemma 5.2 in \cite{DraismadPP1999}). The claim follows by\eqref{Bound.H} and the same arguments for Case a.  Consequently, we complete the proof.
\QED

{\sat\label{Pro_inv}
Let $a\neq0, c, d\in\R$ and $\alpha\in\R, \rho, \eta<0$. If $ f(x)= a x^{\alpha}(1+c x^{\rho} +d x^{2\rho}(1+o(1))), x\to\IF$, then $f\in3\RV_{\alpha, \rho,\rho}$ with auxiliary functions $A, B$ given by
\BQNY
A(t) =\frac{\rho c t^\rho}{1+ct^\rho}, \quad B(t) = \frac{2d}ct^\rho.
\EQNY
Further for \cL{$\alpha\neq0$} as $t\to f(\IF):=\lim_{x\to\IF}f(x)$
\BQNY
f^\leftarrow(t)=\fracl ta^{1/\alpha}\left(1-\frac c\alpha \fracl ta^{\rho/\alpha} +\left(\frac{c^2}{2\alpha^2} (1+\alpha+2\rho) - \frac d\alpha\right)\fracl ta^{2\rho/\alpha}(1+o(1))\right).
\EQNY
}
\proofprop{Pro_inv} The first claim follows by the definition of third-order regularly varying functions. \\
We only present the proof of the second claim for $\alpha>0$ since the case $\alpha<0$ follows by the similar arguments. Since $f\in 2\RV_{\alpha, \rho}$ with auxiliary function $A$, we have by Proposition 2.5 in \cite{MaoH2012} that
\BQN\label{Secon-inverse}
(f^\leftarrow(t))^\rho& =& \fracl ta^{\rho/\alpha}\left(1-\frac c\alpha \fracl ta^{\rho/\alpha} (1+o(1)) \right)^\rho \notag\\
& =& \fracl ta^{\rho/\alpha}\left(1-\frac{\rho c}\alpha \fracl ta^{\rho/\alpha} (1+o(1)) \right), \quad t\to f(\IF). 
\EQN
By Theorem 1.5.12 in \cite{BinghamGT1987} we have $f(f^\leftarrow(t))\sim t, t\to f(\IF)$. 
Consequently
\BQNY
f^\leftarrow(t) &=& \fracl ta^{1/\alpha}\left(1+c(f^\leftarrow(t))^\rho + d(f^\leftarrow(t))^{2\rho}(1+o(1)) \right)^{-1/\alpha}\\
&=&  \fracl ta^{1/\alpha}\left(1-\frac c\alpha(f^\leftarrow(t))^\rho + \left(\frac{(1+\alpha)c^2}{2\alpha^2}-\frac d\alpha\right)(f^\leftarrow(t))^{2\rho}(1+o(1)) \right)
\EQNY
which together with \eqref{Secon-inverse} implies the desired result. 
 \QED

 \bibliographystyle{plain}
 \bibliography{ThirdD}

\begin{thebibliography}{10}

\bibitem{BarbeM2004}
Ph. Barbe and W.~P. McCormick.
\newblock Asymptotic expansions for infinite weighted convolutions of heavy
  tail distributions and applications.
\newblock {\em http: arXiv:math/0412537}, 2004.

\bibitem{BelliniB2014}
F.~Bellini and E.D. Bernardino.
\newblock Risk management with {E}xpectiles.
\newblock Preprint, 2015.

\bibitem{BelliniR2008}
F.~Bellini and Rosazza~Gianin E.
\newblock On {H}aezendonck risk measures.
\newblock {\em Journal of Banking and Finance}, 32:986--994, 2008.

\bibitem{BelliniRosazza2012}
F.~Bellini and Rosazza~Gianin E.
\newblock Haezendonck-{G}oovaerts risk measures and {O}rlicz quantiles.
\newblock {\em Insurance Math. Econom.}, 51(1):107--114, 2012.

\bibitem{BelliniKMR2014}
F.~Bellini, B.~Klar, A.~M{\"u}ller, and Rosazza~Gianin E.
\newblock Generalized quantiles as risk measures.
\newblock {\em Insurance Math. Econom.}, 54:41--48, 2014.

\bibitem{BinghamGT1987}
N.H. Bingham, C.M. Goldie, and J.L. Teugels.
\newblock {\em Regular variation}, volume~27 of {\em Encyclopedia of
  Mathematics and its Applications}.
\newblock Cambridge University Press, Cambridge, 1987.

\bibitem{CaeiroG2008}
F.~Caeiro and M.~I. Gomes.
\newblock Minimum-variance reduced-bias tail index and high quantile
  estimation.
\newblock {\em REVSTAT}, 6(1):1--20, 2008.

\bibitem{CaideHZ2013}
J.~Cai, L.~de~Haan, and C.~Zhou.
\newblock Bias correction in extreme value statistics with index around zero.
\newblock {\em Extremes}, 16(2):173--201, 2013.

\bibitem{CaiW2014}
J.~Cai and C.~Weng.
\newblock Optimal reinsurance with expectile.
\newblock {\em Scand. Actuar. J.},
  (http://dx.doi.org/10.1080/03461238.2014.994025), 2014.

\bibitem{deHaanF2006}
L.~de~Haan and A.~Ferreira.
\newblock {\em Extreme value theory}.
\newblock Springer Series in Operations Research and Financial Engineering.
  Springer, New York, 2006.
\newblock An introduction.

\bibitem{deHaanP1997}
L.~de~Haan and L.~Peng.
\newblock Rates of convergence for bivariate extremes.
\newblock {\em J. Multivariate Anal.}, 61(2):195--230, 1997.

\bibitem{deHaanS1996}
L.~de~Haan and U.~Stadtm{\"u}ller.
\newblock Generalized regular variation of second order.
\newblock {\em J. Austral. Math. Soc. Ser. A}, 61(3):381--395, 1996.

\bibitem{DraismadPP1999}
G.~Draisma, L.~de~Haan, L.~Peng, and T.T. Pereira.
\newblock A bootstrap-based method to achieve optimality in estimating the
  extreme-value index.
\newblock {\em Extremes}, 2(4):367--404 (2000), 1999.

\bibitem{EmbrechtsKM1997}
P.~Embrechts, C.~Kl{\"u}ppelberg, and T.~Mikosch.
\newblock {\em Modelling extremal events}, volume~33 of {\em Applications of
  Mathematics (New York)}.
\newblock Springer-Verlag, Berlin, 1997.
\newblock For insurance and finance.

\bibitem{FragaAlvesdL2006}
I.~Fraga~Alves, L.~de~Haan, and T.~Lin.
\newblock Third order extended regular variation.
\newblock {\em Publ. Inst. Math. (Beograd) (N.S.)}, 80(94):109--120, 2006.

\bibitem{Gelukd1987}
J.L. Geluk and L.~de~Haan.
\newblock {\em Regular variation, extensions and {T}auberian theorems},
  volume~40 of {\em CWI Tract}.
\newblock Stichting Mathematisch Centrum, Centrum voor Wiskunde en Informatica,
  Amsterdam, 1987.

\bibitem{Gneiting2011}
T.~Gneiting.
\newblock Making and evaluating point forecasts.
\newblock {\em J. Amer. Statist. Assoc.}, 106(494):746--762, 2011.

\bibitem{HaezendonckG1982}
J.~Haezendonck and M.~Goovaerts.
\newblock A new premium calculation principle based on {O}rlicz norms.
\newblock {\em Insurance Math. Econom.}, 1(1):41--53, 1982.

\bibitem{HashorvaLP14}
E.~Hashorva, C.~Ling, and Z.~Peng.
\newblock Second order tail asymptotics of deflated risks.
\newblock {\em Insurance Math. Econom.}, 56:88--101.

\bibitem{HashorvaPT2010}
E.~Hashorva, A.~G. Pakes, and Q.~Tang.
\newblock Asymptotics of random contractions.
\newblock {\em Insurance Math. Econom.}, 47(3):405--414, 2010.

\bibitem{HashorvaP2010}
E.~Hashorva and A.G. Pakes.
\newblock Tail asymptotics under beta random scaling.
\newblock {\em J. Math. Anal. Appl.}, 372(2):496--514, 2010.

\bibitem{LiPX2011}
D.~Li, L.~Peng, and X.~Xu.
\newblock Bias reduction for endpoint estimation.
\newblock {\em Extremes}, 14(4):393--412, 2011.

\bibitem{MaoBook2013}
T.~Mao.
\newblock Second-order conditions of regular variation and {D}rees-type
  inequalities.
\newblock In {\em Stochastic orders in reliability and risk}, volume 208 of
  {\em Lecture Notes in Statist.}, pages 313--330. Springer, New York, 2013.

\bibitem{MaoH2012}
T.~Mao and T.~Hu.
\newblock Second-order properties of the {H}aezendonck-{G}oovaerts risk measure
  for extreme risks.
\newblock {\em Insurance Math. Econom.}, 51(2):333--343, 2012.

\bibitem{MaoHN2015}
T.~Mao, T.~Hu, and K.~Ng.
\newblock Asymptotics of generalized quantitles and {E}xpectiles for extreme
  risks.
\newblock {\em Probability in the Engineering and Informational Sciences}, to
  appear, 2015.

\bibitem{Neves2009}
C.~Neves.
\newblock From extended regular variation to regular variation with application
  in extreme value statistics.
\newblock {\em J. Math. Anal. Appl.}, 355(1):216--230, 2009.

\bibitem{NeweyP1987}
W.K. Newey and J.L. Powell.
\newblock Asymmetric least squares estimation and testing.
\newblock {\em Econometrica}, 55(4):819--847, 1987.

\bibitem{OliveriraGF2006}
O.A. Oliveira, M.I. Gomes, and M.I. Fraga~Alves.
\newblock Improvements in the estimation of a heavy tail.
\newblock {\em REVSTAT}, 4(2):81--109, 2006.

\bibitem{PakesN2007}
A.G. Pakes and J.~Navarro.
\newblock Distributional characterizations through scaling relations.
\newblock {\em Aust. N. Z. J. Stat.}, 49(2):115--135, 2007.

\bibitem{PengNL2010}
Z.~Peng, S.~Nadarajah, and F.~Lin.
\newblock Convergence rate of extremes for the general error distribution.
\newblock {\em J. Appl. Probab.}, 47(3):668--679, 2010.

\bibitem{ReissT2007}
R.-D. Reiss and M.~Thomas.
\newblock {\em Statistical analysis of extreme values with applications to
  insurance, finance, hydrology and other fields}.
\newblock Birkh\"auser Verlag, Basel, 2007.

\bibitem{Resnick2007}
S.I. Resnick.
\newblock {\em Heavy-tail phenomena}.
\newblock Springer Series in Operations Research and Financial Engineering.
  Springer, New York, 2007.
\newblock Probabilistic and statistical modeling.

\bibitem{TangY2012}
Q.~Tang and F.~Yang.
\newblock On the {H}aezendonck-{G}oovaerts risk measure for extreme risks.
\newblock {\em Insurance Math. Econom.}, 50(1):217--227, 2012.

\bibitem{TangY2014}
Q.~Tang and F.~Yang.
\newblock Extreme value analysis of the {H}aezendonck-{G}oovaerts risk measure
  with a general {Y}oung function.
\newblock {\em Insurance Math. Econom.}, 59:311--320, 2014.

\bibitem{WangZ2014}
R.~Wang and J.F. Ziegel.
\newblock Distortion risk measures and elicitability.
\newblock Papers, arXiv:1405.3769, 2014.

\bibitem{WangC2006}
X.~Wang and S.~Cheng.
\newblock General regular variation of {$n$}-th order and the 2nd order
  {E}dgeworth expansion of the extreme value distribution. {II}.
\newblock {\em Acta Math. Sin. (Engl. Ser.)}, 22(1):27--40, 2006.

\bibitem{Ziegel2014}
J.F. Ziegel.
\newblock Coherence and {E}licitability.
\newblock {\em Mathematical Finance}, (494):1--18, 2014.

\end{thebibliography}
\end{document}